\documentclass[aps,prx,preprint,onecolumn,citeautoscript,superscriptaddress,footinbib,
eqsecnum]{revtex4-1}  
\usepackage{amsmath,amssymb,bm,bbm} 
\usepackage{amsthm}
\usepackage{graphicx}  
\usepackage{color} 
\usepackage[dvipsnames]{xcolor}
\usepackage[papersize={8.5in,11in}]{geometry}
\usepackage[colorlinks=true]{hyperref}  
\usepackage[section]{placeins}
\hypersetup{
    bookmarks=true,         
    unicode=false,          
    pdftoolbar=true,        
    pdfmenubar=true,        
    pdffitwindow=false,     
    pdfstartview={FitH},    
    pdftitle={Z2 fractionalized phases of a solvable, disordered, $t$-$J$ model},    
    pdfauthor={Wenbo Fu, Yingfei Gu, Subir Sachdev, Grigory Tarnopolsky},     
    pdfsubject={},   
    pdfcreator={},   
    pdfproducer={}, 
    pdfkeywords={} {} {}, 
    pdfnewwindow=true,      
    colorlinks=true,       
    linkcolor=magenta, 
    citecolor=blue,        
    filecolor=magenta,      
    urlcolor=blue           
} 

\usepackage{tikz}
\usepackage{tikz-cd}
\usetikzlibrary{arrows}
\usetikzlibrary{intersections}
\usetikzlibrary{shapes.geometric}
\usetikzlibrary{decorations.pathmorphing, patterns,shapes}

\renewcommand{\bar}{\overline}
\renewcommand{\tilde}{\widetilde}

\newcommand{\nn}{\nonumber}

\newcommand{\const}{\operatorname{const}}

\newcommand{\ZZ}{\mathbb{Z}}

\usepackage{amsfonts}
\usepackage{upgreek}
\usepackage{slashed}
\usepackage{latexsym}
\usepackage{todonotes}

\newcommand{\beq}{\begin{equation}}
\newcommand{\eeq}{\end{equation}}
\def\bea{\begin{eqnarray}}
\def\eea{\end{eqnarray}}
\newcommand{\sgn}{\operatorname{sgn}}

\begin{document}

\title{$\mathbb{Z}_2$ fractionalized phases of a solvable, disordered, $t$-$J$ model}

\author{Wenbo Fu}
\affiliation{Department of Physics, Harvard University, Cambridge MA 02138, USA}

\author{Yingfei Gu}
\affiliation{Department of Physics, Harvard University, Cambridge MA 02138, USA}

\author{Subir Sachdev}
\affiliation{Department of Physics, Harvard University, Cambridge MA 02138, USA}
\affiliation{Perimeter Institute for Theoretical Physics, Waterloo, Ontario, Canada N2L 2Y5}

\author{Grigory Tarnopolsky}
\affiliation{Department of Physics, Harvard University, Cambridge MA 02138, USA}

\date{\today
\\
\vspace{0.4in}}

\begin{abstract}
We describe the phases of a solvable $t$-$J$ model of electrons with infinite-range, and random, hopping and exchange interactions, similar to those in the Sachdev-Ye-Kitaev models. The electron fractionalizes, as in an 
`orthogonal metal', into a fermion $f$ which carries both the electron spin and charge, 
and a boson $\phi$. Both $f$ and $\phi$ carry emergent $\mathbb{Z}_2$ gauge charges.
The model has a phase in which the $\phi$ bosons are gapped, and the $f$ fermions are gapless and critical, and so the electron spectral function is gapped. This phase can be considered as a toy model for the underdoped cuprates.
The model also has an extended, critical, `quasi-Higgs' phase 
where both $\phi$ and $f$ are gapless,
and the electron operator $\sim f \phi$ has a Fermi liquid-like $1/\tau$ propagator in imaginary time, $\tau$. So while the electron spectral function has a Fermi liquid form, other properties are controlled by $\mathbb{Z}_2$ fractionalization and the anomalous exponents of the $f$ and $\phi$ excitations.
This `quasi-Higgs' phase is proposed  as a toy model of the overdoped cuprates. We also describe the critical state separating these two phases.
\end{abstract}

\maketitle

\section{Introduction}

One of the main outstanding puzzles in the study of the cuprate superconductors is the nature of the transformation 
in the electronic state near optimal doping. There are numerous experimental indications that the underlying electronic
state changes from a Mott-like state with a small density of carriers at low doping, to a Fermi liquid-like state with a large
density of carriers at high doping. The most recent indication of this transformation is in the doping dependence of Hall coefficient \cite{badoux2016}. It is also becoming clear that this phenomenology cannot be described solely in terms of a conventional
symmetry-breaking phase transition in the Landau framework: despite much experimental effort, no suitable order parameter with sufficient
strength has been found near optimal doping. Furthermore such order parameters are also sensitive to quenched disorder, while the 
cuprate transition appears quite robust to varying degrees of disorder {\em e.g.\/} the transformation in the electronic state is seen
in STM experiments in both the `2212' and `2201' series of compounds \cite{He13,Fujita14}. 

The most promising route therefore appears to lie in investigating non-Landau transitions which have a `topological' character. 
Moreover, we need to understand such transitions in the presence of finite density fermionic matter, and also with quenched randomness. 
There are no known theories of quantum phase transitions under such conditions. Solvable examples in simple limits would clearly be valuable.

In this paper we propose a solvable $t$-$J$ model of electrons which exhibits a phase transition under such conditions. Both phases of our model are deconfined, possessing gapless fermionic excitations, $f$, which carry $\mathbb{Z}_2$ gauge charges. Our model also possesses a bosonic ``Higgs'' field $\phi$, carrying $\mathbb{Z}_2$ gauge charges,
and the electron is a composite of $f$ and $\phi$.
The Higgs field is gapped in one of the phases, and so is the electron: this phase can be considered as a toy model for the underdoped cuprates. The other phase has power-law correlations of the Higgs field: so it is not quite a Higgs/confining phase of the $\mathbb{Z}_2$ gauge theory, but a novel `quasi-Higgs' phase with slowly decaying correlations of the Higgs field. 
The electron operator in this quasi-Higgs phase has a leading $1/\tau$ decay
in imaginary time, as in a Fermi liquid. We propose this phase as a toy model
for the overdoped cuprates. 

Our model is a 0+1 dimensional quantum theory, in the class of the Sachdev-Ye-Kitaev (SYK) models \cite{SY92,kitaev2015talk}. Although these models
do not have any spatial structure, they exhibit a `local criticality' which is  
 interesting for a number of physical questions:
\begin{itemize}
\item The SYK models are the simplest solvable models without quasiparticle excitations. So they can be used as fully quantum building blocks
for theories of strange metals \cite{PG98,Gu2017local,Davison17,Gu17,Balents17,Patel18,DC18}.
\item The SYK models exhibit many-body chaos \cite{kitaev2015talk,Maldacena2016}, and saturate the lower bound on the Lyapunov time of large-$N$ model to reach chaos \cite{Maldacena2016a}. So they are ``the most chaotic'' quantum many-body systems. The presence of maximal chaos is linked to the absence of quasiparticle excitations,  and the proposed \cite{ssbook} lower bound of order $\hbar /(k_B T)$  on a `dephasing time'. 
\item Related to their chaos, the SYK models exhibit \cite{Sonner17} eigenstate thermalization (ETH) \cite{Deutsch91,Srednicki94}, and yet many aspects are exactly solvable.
\item The SYK models are dual to gravitational theories in $1+1$ dimensions which have a black hole horizon. The connection between the SYK models and black holes with a near-horizon AdS$_2$ geometry was proposed in Refs.~\cite{SS10,SS10b}, and made much sharper in Refs.~\cite{kitaev2015talk,nearlyads2,kitaev2017}. It has been used to examine aspects of the black hole information problem \cite{Maldacena2017}. 
\end{itemize}

We model the underdoped state
of the cuprate superconductors as a deconfined phase of a $\mathbb{Z}_2$ gauge theory \cite{SSDC16}. 
The case which we have found to be most amenable
to a SYK-like description is to represent the deconfined phase as an `orthogonal metal' \cite{OM12,OM10}.
In this description, the electron operator $c_{i \alpha}$ ($i$ is a site index, and $\alpha$ is a spin label) 
fractionalizes into an `orthogonal fermion', $f_{i \alpha}$, which carries both the spin and charge of the electron, and an Ising variable $\sigma^z_i$:
\beq
c_{i \alpha} = \sigma^z_i \, f_{i \alpha} \,.
\eeq
Note that this decomposition is invariant under the $\mathbb{Z}_2$ gauge transformation 
\beq
\sigma^z_i \rightarrow \eta_i \sigma^z_i \quad , \quad f_{i \alpha} \rightarrow \eta_i f_{i \alpha}, 
\eeq
where $\eta_i = \pm 1$.
We can then set up a $t$-$J$ model for these degrees of freedom, with a Hamiltonian like
\beq
\mathcal{H}_\sigma = - \sum_{i,j} t_{ij} \sigma^z_i \sigma^z_j \, f_{i \alpha}^\dagger f_{i \alpha} + 
\sum_{i > j,\alpha\beta} J_{ij} f_{i \alpha}^\dagger f_{i \beta} f_{j \beta}^\dagger f_{j \alpha} - g \sum_i \sigma^x_i\,.
\eeq
At large $g$, the value of $\sigma^z$ will rapidly average to zero, and only the $J_{ij}$ term will be active: so we expect a fractionalized
orthogonal metal state in which the $\sigma^z$ excitations are gapped, and the orthogonal fermions $f$ are deconfined.
In contrast, at small $g$, the $\sigma^z$ can condense and then $\mathbb{Z}_2$ charges are confined: this would
be a conventional state in which $c \sim f$. Indeed, a similar transition has appeared in a recent Monte Carlo study on the square lattice at half-filling, between an orthogonal semi-metal and a confining superconductor or a confining antiferromagnet \cite{Gazit17,Assaad16,Gazit18}.
However, as we noted above, the specific model we shall study here only has a gapless, `almost confining', quasi-Higgs phase.

\begin{figure}
\begin{center}
\includegraphics[width=2.8in]{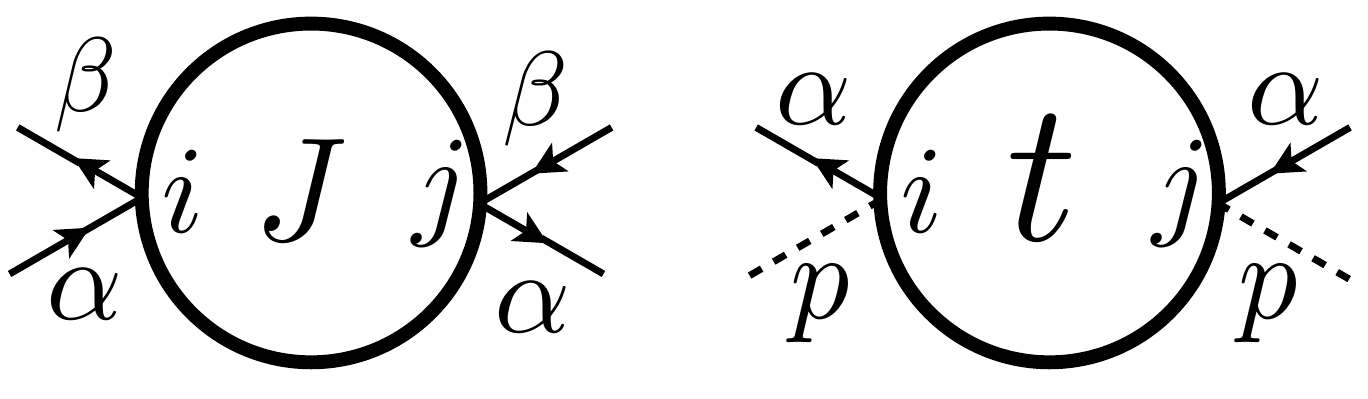}
\end{center}
\caption{The interaction vertices in $\mathcal{L}$. The full lines are $f$ fermions, and the dashed lines are $\phi$ bosons.}
\label{fig:vertices}
\end{figure}

The model $\mathcal{H}_{\sigma}$ is not directly amenable to a SYK-like large $N$ limit. However, it does become so when we promote
the $\mathbb{Z}_2$ Ising spin to an O($M'$) quantum rotor \cite{YSR93,RSY94}, $\phi_p$, $p = 1 \ldots M'$ which obeys the constraint
\beq
\sum_{p=1}^{M'} \phi_{ip}^2 = M' \,.
\label{unit}
\eeq
As in Ref.~\cite{YSR93,RSY94}, we expect that this promotion from Ising to large $M'$ rotors does
not modify the universal critical properties.
To obtain a suitable large $M$ limit, we also promote the spin index
$\alpha = 1 \ldots M$ to a SU($M$) spin index (as in Ref.~\cite{SY92}).
For this purpose, we introduce an orbital index, $p$, and fractionalize the electron as
\beq
c_{i p \alpha} = \phi_{ip} f_{i \alpha} \,, \label{defc}
\eeq
so that 
\beq
\phi_{i p} \rightarrow \eta_i \phi_{ip}
\eeq
under the $\mathbb{Z}_2$ gauge transformation.
Then we obtain the final Lagrangian of the $t$-$J$ model to be solved in this paper: 
\bea
\mathcal{L} &=& \frac{1}{2g}\sum_{i,p}\left( \partial_\tau \phi_{ip} \right)^2 + \sum_{i,\alpha} f_{i \alpha}^\dagger \left( \frac{\partial}{\partial \tau} - \mu \right) f_{i \alpha} \nn \\
&~&~+ \frac{1}{\sqrt{NM}} \sum_{i,j,p,\alpha} t_{ij} \phi_{ip} \phi_{jp} f_{i \alpha}^\dagger f_{j \alpha}  +
\frac{1}{\sqrt{NM}} \sum_{i > j,\alpha\beta} J_{ij} f_{i \alpha}^\dagger f_{i \beta} f_{j \beta}^\dagger f_{j \alpha} \,,
\eea
where the site indices $i,j=1 \ldots N$.
With $t_{ij}$ and $J_{ij}$ independent random numbers with zero mean, 
we will show that this Lagrangian is solvable in the limit of large number of sites, $N$, followed by the limit of large $M$ and $M'$
at fixed 
\beq
k \equiv \frac{M'}{M} \,.
\eeq
For our future diagrammatic analysis, we represent the interaction vertices in $\mathcal{L}$ in Fig.~\ref{fig:vertices}.

\section{Large $N$ limit}
\label{sec:largeN}

To take the large $N$ limit, we average over $t_{ij}$ and $J_{ij}$, with $\overline{|t_{ij}|^2} = t^2/2$ and
$\overline{J_{ij}^2} = J^2$. As usual, everything reduces to a single site
problem, with the fields carrying replica indices. However, for simplicity, we drop the replica indices. 
Then the single-site Lagrangian is 
\bea
\mathcal{L} &=& \frac{1}{2g}\sum_{p}\left( \partial_\tau \phi_{p} \right)^2 
+ i \lambda \left(\sum_p \phi_p^2 - M' \right)
+ \sum_{\alpha} f_{\alpha}^\dagger \left( \frac{\partial}{\partial \tau} - \mu \right) f_{\alpha} \nn \\
&~&~-\frac{t^2}{M} \sum_{p,\alpha} \int_{0}^{1/T} d \tau d \tau' R^\ast (\tau - \tau') 
\phi_{p} (\tau) \phi_{p} (\tau') f_{\alpha}^\dagger (\tau ) 
f_{\alpha} (\tau ')  \nn \\ &~&~-
\frac{J^2}{2M} \sum_{\alpha,\beta} \int_{0}^{1/T} d \tau d \tau'  Q(\tau-\tau') f_{\alpha}^\dagger (\tau) f_{\beta} (\tau) f_{\beta}^\dagger (\tau') f_{\alpha} (\tau') \,,
\eea
where $T$ is the temperature and  $\lambda$ is the Lagrange multiplier imposing Eq.~(\ref{unit}). More precisely, as in Ref.~\cite{SY92}, decoupling the large $N$ path integral introduces fields analogous to $R$ and $Q$ which are off-diagonal in the SU($M$) and O($M'$) indices. We have assumed above that the large $N$ limit is dominated by the saddle point in which these fields are SU($M$) and O($M'$) diagonal. This requires that the large $N$ limit is taken {\it before\/} the large $M$  and $M'$ limits. This procedure supplements the Lagrangian with the self-consistency conditions
\bea
R(\tau - \tau') &=& -\frac{1}{M M'} \sum_{p,\alpha} \left\langle \phi_{p} (\tau) \phi_{p} (\tau') f_{\alpha}^\dagger (\tau ) 
f_{\alpha} (\tau ') \right\rangle \nn \\
Q(\tau - \tau') &=& \frac{1}{M^2} \sum_{\alpha,\beta} \left\langle f_{\alpha}^\dagger (\tau) f_{\beta} (\tau) f_{\beta}^\dagger (\tau') f_{\alpha} (\tau') \right\rangle\,.
\eea
It is convenient to rescale $\phi_p \rightarrow \sqrt{g} \phi_p$ so that the Lagrangian becomes
\bea
\mathcal{L} &=& \frac{1}{2}\sum_{p}\left( \partial_\tau \phi_{p} \right)^2 
+ i \lambda \left(\sum_p \phi_p^2 - \frac{M'}{g} \right)
+ \sum_{\alpha} f_{\alpha}^\dagger \left( \frac{\partial}{\partial \tau} - \mu \right) f_{\alpha} \nn \\
&~&~-\frac{\tilde{t}\,^2}{M} \sum_{p,\alpha} \int_{0}^{1/T} d \tau d \tau' R^\ast (\tau - \tau') 
\phi_{p} (\tau) \phi_{p} (\tau') f_{\alpha}^\dagger (\tau ) 
f_{\alpha} (\tau ')   \nn \\ &~&~-
\frac{J^2}{2M} \sum_{\alpha,\beta} \int_{0}^{1/T} d \tau d \tau'  Q(\tau-\tau') f_{\alpha}^\dagger (\tau) f_{\beta} (\tau) f_{\beta}^\dagger (\tau') f_{\alpha} (\tau') \,,
\eea
where $\tilde{t} = t g$. 

\begin{figure}
\begin{center}
\includegraphics[width=3.5in]{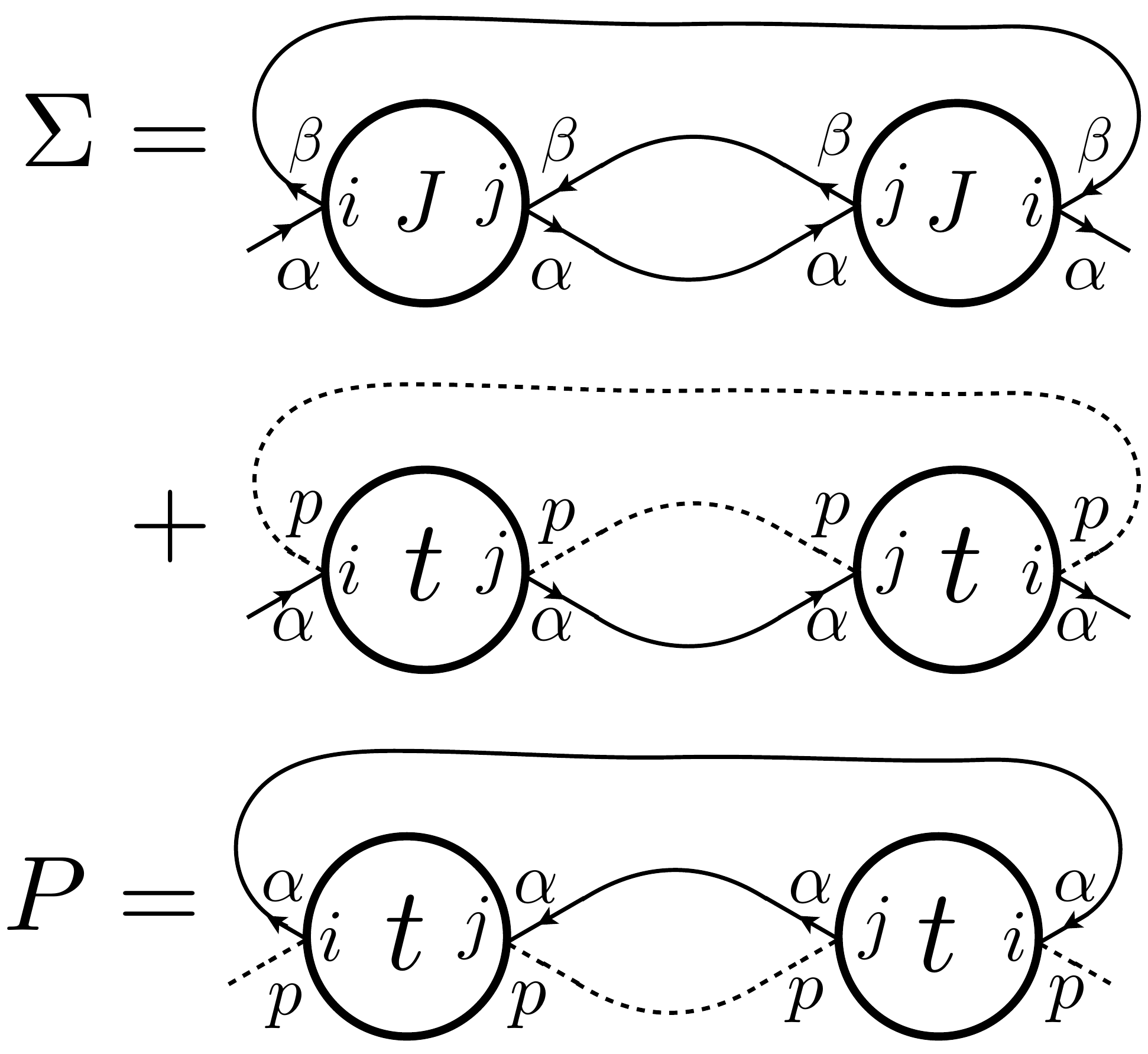}
\end{center}
\caption{Feynman diagrams for the self energies of the fermions and bosons in Eqs.~(\ref{feqns}) and (\ref{beqns}). The sum over the internal site index $j$ yields a factor of $N$. The sums over the loops of
SU($M$) spin indices $\alpha$  and $\beta$ yield factors of $M$. The sum over the loop of the O($M'$) index $p$ yields a factor $M'$.}
\label{fig:selfenergy}
\end{figure}

Next we take the large $M$ and $M'$ limit at fixed $k=M/M'$. Note that the large $N$ limit
has already been taken. By this sequence of limits we obtain for the fermion Green's function, $G$, and the $\phi$ correlator $\chi$
\bea
G(i \omega_n) &=& \frac{1}{i \omega_n + \mu - \Sigma (i \omega_n)} \quad, \quad \Sigma (\tau) = - J^2 G^2 (\tau) G(-\tau)
+ k \, \tilde{t}\,^2 G(\tau) \chi^2 (\tau) \label{feqns} \\
\chi ( i \omega_n) &=& \frac{1}{\omega_n^2 + \chi_0^{-1} -  P (i \omega_n) + P(i\omega_n=0)} \quad, \quad P (\tau) = - 2\,\tilde{t}\,^2 G(\tau) G(-\tau) \chi (\tau) \label{beqns}
\eea
where 
\beq 
i\lambda = \chi_0^{-1} + P(i \omega_n = 0)
\label{lambda0}
\eeq
is the saddle point value of $i \lambda$. 
Note that we have introduced notation so that
\beq
\chi (i\omega_n = 0) \equiv \chi_0 \,,
\label{static}
\eeq
is the static $\phi$ susceptibility. Formally, the value of $\chi_0$ is to be determined by solving the constraint equation Eq.~(\ref{unit}):
\beq
T \sum_{\omega_n} \chi (i \omega_n) = \frac{1}{g}\,. \label{gval}
\eeq
In practice, we will treat the value of $\chi_0$ as a parameter that can be tuned to access all the regions of the phase diagram, and use
Eq.~(\ref{gval}) to determine the value of $g$. This is convenient because the coupling $g$ does not appear in any of the 
other saddle-point equations (after our definition of $\tilde{t}$). Finally, the results as a function of $\chi_0$ will be recast as functions of $g$.
We note that the large $N$ equations in Eqs.~(\ref{feqns}) and (\ref{beqns}) can also be derived
diagrammatically, as illustrated in Fig.~\ref{fig:selfenergy}.

Coupled equations of Green's functions of bosons and fermions have been considered previously in a supersymmetric model \cite{FGMS17}, but the present equations have a different structure. The supersymmetric model has a single boson field coupling to fermion composites, while $\mathbb{Z}_2$ gauge invariance of our model requires
that pairs of bosons couple to fermions.

\section{Gapless solutions}
\label{sec:gapless}

First, we search for solutions of Eqs.~(\ref{feqns}), (\ref{beqns}), and (\ref{gval}) in which both the fermions and the bosons are gapless.
In our initial analysis, we will work on the imaginary frequency axis at $T=0$ (see Appendix~\ref{app:spectral} for definitions of spectral functions). The extension to $T>0$ appears in Section~\ref{sec:gaplessT}.

For the gapless solutions, we make the ansatzes valid as $\tau \rightarrow \infty$ at $T=0$
\bea
G(\tau) &=& - \mbox{sgn} (\tau) \frac{F}{(J |\tau|)^{2 \Delta_f}} \nn \\
 \chi (\tau) &=& \frac{C/J}{(J|\tau|)^{2 \Delta_b}} \,, \label{chicrit}
\eea
where $F>0$ and $C>0$, and they are both dimensionless.
The Fourier transforms at $T=0$ are
\bea
G(i \omega) &=& - 2i \, \mbox{sgn} (\omega)\frac{F/J^{2 \Delta_f}}{|\omega|^{1 - 2 \Delta_f}} \cos (\pi \Delta_f) \Gamma( 1- 2 \Delta_f) \nn \\
\chi (i \omega) &=&  2 \frac{C/J^{2 \Delta_b+1}}{|\omega|^{1 - 2 \Delta_b}} \sin (\pi \Delta_b) \Gamma( 1- 2 \Delta_b)\,.
\label{g2}
\eea
From Eq.~(\ref{feqns}) and (\ref{beqns}), the self energies are 
\bea
\Sigma (\tau ) &=& - \mbox{sgn} (\tau) \left(\frac{J^2 F^3}{(J|\tau|)^{6 \Delta_f}} + \frac{ k \,\tilde{t}\,^2 F C^2/J^2}{(J|\tau|)^{2 \Delta_f + 4 \Delta_b}} \right) \nn \\
P(\tau) &=&  \frac{2\,\tilde{t}\,^2 F^2 C/J}{(J|\tau|)^{4 \Delta_f + 2 \Delta_b}}\,,
\eea
and their Fourier transforms are 
\bea
\Sigma (i \omega ) &=& - 2 i \, \mbox{sgn} (\omega) \left(\frac{J^{2-6 \Delta_f} F^3}{|\omega|^{1-6 \Delta_f}}
\cos(3 \pi \Delta_f) \Gamma(1 - 6 \Delta_f) \right. \nn \\
&~&~~~~~~~~\left. 
 + \frac{ k (\tilde{t}/J^2)^2 J^{2- 2 \Delta_f - 4 \Delta_b} F C^2}{|\omega|^{1-2 \Delta_f - 4 \Delta_b}} 
 \cos(\pi (\Delta_f + 2 \Delta_b)) \Gamma(1 - 2 \Delta_f - 4 \Delta_b)
 \right) \nn \\
P(i \omega) &=&  4 \frac{(\tilde{t}/J^2)^2 J^{3- 4 \Delta_f - 2 \Delta_b} F^2 C}{|\omega|^{1-4 \Delta_f - 2 \Delta_b}}
\sin(\pi (2 \Delta_f + \Delta_b)) \Gamma(1 - 4\Delta_f - 2 \Delta_b)
\,. \label{se2}
\eea
From Eqns~(\ref{g2}) and (\ref{se2}), and using $G(i \omega) \Sigma (i \omega) = -1$
and $\chi (i \omega) P(i \omega) = -1$ in the limit of low $\omega$, we see that solutions are only possible
when
\beq
\Delta_f + \Delta_b = 1/2 \,. \label{Deltafb}
\eeq
Further examination of the saddle point equations shows that two classes of solutions are possible,  depending upon whether $\Delta_f > 1/4$ 
or $\Delta_f = 1/4$. We will examine these solutions in the following subsections.

\subsection{$\Delta_f > 1/4$}
\label{sec:deltag14}

In this case, the first term in $\Sigma (i \omega)$ in Eq.~(\ref{se2}) is subdominant and can be ignored.
Then the Schwinger-Dyson equations simplify
\bea
 k (\tilde{t}/J^2)^2 F^2 C^2 \frac{4 \pi \cot(\pi \Delta_f)}{2 - 4 \Delta_f} &=& 1 \nn \\
(\tilde{t}/J^2)^2 F^2 C^2  \frac{2 \pi \cot(\pi \Delta_f)}{\Delta_f}&=& 1
\label{eq:FC}
\eea
These equations are consistent only if we choose the scaling dimensions
\beq
\Delta_f = \frac{1}{k + 2} \quad , \quad \Delta_b = \frac{k}{2 (k + 2)} \,.
\label{alphaexp}
\eeq
Note that $\Delta_f > 1/4$ requires $k < 2$. So the exponents are limited to the ranges
\beq
\frac{1}{4} < \Delta_f < \frac{1}{2} \quad , \quad 0 < \Delta_b < \frac{1}{4} \,.
\eeq

The above analysis of the low $\omega$ limit of the saddle point equations does not determine the values of $F$ and $C$ separately, only
the value of their product $C F$. So we expect that the $\Delta_f > 1/4$ solution defines a phase which extends over a range of value of $g$.
Our numerical analysis will confirm that this is indeed the case.

\subsection{$\Delta_f = \Delta_b =  1/4$}
\label{sec:delta14}

Now both terms in $\Sigma$ in Eq.~(\ref{se2}) have the same frequency dependence, and so both contribute to the low $\omega$ limit.
The Schwinger-Dyson equations now become
\bea
 F^4 + k (\tilde{t}/J^2)^2 C^2 F^2 &=& \frac{1}{4 \pi} \nn \\
(\tilde{t}/J^2)^2 C^2 F^2 &=& \frac{1}{8 \pi} \,. \label{e2}
\eea
These can be solved uniquely for both $F>0$ and $C>0$ provided again $k < 2$. 
The existence of unique low $\omega$ solution with these exponents indicates that Eq.~(\ref{gval}) will yield only a particular
value of $g$. We will find that is the case in our numerics, and this solution appears to describe a critical point between our 
$\Delta_f > 1/4$ gapless and gapped phases.

\subsection{Non-zero temperatures}
\label{sec:gaplessT}

It turns out that a $T>0$ conformal extension of the above gapless solutions satisfies the saddle point equations in Eqs.~(\ref{feqns}) and (\ref{beqns})
at $T>0$, just as was noted in Refs.~\cite{PG98,SS10b}.  From Eq.~(\ref{chicrit}), the conformal extension is 
\bea
G(\tau) &=& - \mbox{sgn} (\tau) \frac{F}{J^{2 \Delta_f}} \left( \frac{\pi T}{|\sin(\pi T \tau)|} \right)^{2 \Delta_f} \nn \\
 \chi (\tau) &=& \frac{C}{J^{2 \Delta_b+1}} \left( \frac{\pi T}{|\sin(\pi T \tau)|} \right)^{2 \Delta_b} \,, \label{chiconform}
\eea

But, we also have to verify that the Eq.~(\ref{gval}) yields the same value of $g$ as at $T=0$. The frequency
summation in Eq.~(\ref{gval}) is dominated by high energies, and we don't expect significant change in the spectral weight at
such frequencies at a small $T>0$. So we need only examine the low frequencies in Eq.~(\ref{gval}), in which case we can use
the conformal solution. To focus on low frequencies, we subtract Eq.~(\ref{gval}) between its $T=0$ and $T>0$ values, and regulate
the higher frequencies by inserting a point-splitting $\tau$. Then the requirement that the value of $g$ is the same at $T=0$
and in the conformal solution is
\beq
\lim_{\tau \rightarrow 0} \left[ \chi(\tau, T) - \chi (\tau, T=0) \right] = 0 \label{gvalT}
\eeq
It is now easy to verify that Eq.~(\ref{chiconform}) does indeed satisfy Eq.~(\ref{gvalT}).

Taking the Fourier transform of Eq.~(\ref{chiconform}), we have the low $\omega$ gapless solution as a function of $\omega$ and $T$
\begin{eqnarray}
G (i\omega_n) &=& \left[-  i  \frac{F  \Pi (2 \Delta_f)}{J^{2 \Delta_f}} \right]  \frac{ T^{2 \Delta_f -1} \,
 \Gamma \left( \displaystyle  \Delta_f  + \frac{\omega_n}{2 \pi T} \right)}{
 \Gamma \left( \displaystyle 1 - \Delta_f + \frac{\omega_n}{2 \pi T} \right)  } 
 \nonumber \\
\chi (i\omega_n) &=& \left[   \frac{C  \widetilde{\Pi} (2 \Delta_b)}{J^{2 \Delta_b+1}} \right]  \frac{ T^{2 \Delta_b -1} \,
 \Gamma \left( \displaystyle  \Delta_b  + \frac{\omega_n}{2 \pi T} \right)}{
 \Gamma \left( \displaystyle 1 - \Delta_b + \frac{\omega_n}{2 \pi T} \right)  } 
\end{eqnarray}
where 
\begin{equation}
\Pi (s) \equiv \pi^{s-1 } 2^s  \cos \left(\displaystyle  \frac{\pi s}{2} \right) \Gamma (1-s) \quad , \quad \widetilde{\Pi} (s) \equiv \pi^{s-1 } 2^s  \sin \left(\displaystyle  \frac{\pi s}{2} \right) \Gamma (1-s)
\end{equation}
From Eq.~(\ref{static}), we therefore obtain the $T$-dependence of the static susceptibility
\beq
\chi_0 = \left[   \frac{C  \widetilde{\Pi} (2 \Delta_b) \Gamma ( \Delta_b )}{J^{2 \Delta_b+1} \Gamma  ( 1 - \Delta_b ) } \right]   T^{2 \Delta_b -1} \,.
\label{chi0c}
\eeq
The susceptibility must have this $T$ dependence at low $T$ to keep $g$ fixed while $T$ varies.

\section{Gapped boson solution}
\label{sec:gapped}

Now we search for a possible solution of Eqs.~(\ref{feqns}) and (\ref{beqns}) 
with a gap in the boson spectrum at $T=0$. 
With an energy gap, $m$, from Eq.~(\ref{ab1}) we can conclude that the boson Green's function decays exponentially at long times.
So we write
\beq
\chi (\tau) = \frac{B/J}{( J |\tau|)^\gamma} e^{- m |\tau|} \quad, \quad |\tau| \gg 1/m ,~ T=0\,, \label{bc3}
\eeq
parameterized by the gap $m$, the exponent $\gamma$ and the dimensionless prefactor $B$. From the spectral analysis in Appendix~\ref{app:spectral}
we conclude that the boson Green's function $\chi(z)$ has branch cuts in the complex frequency plane at $z = \pm m$. At $z=m$, the singular (non-analytic) 
part of $\chi (z)$ is
\beq
\chi_{\rm sing} (z) = \frac{\pi B}{J^2 \Gamma(\gamma)} \frac{i J^{1-\gamma}}{(z-m)^{1-\gamma}} \quad , \quad z \sim m \label{bc1}
\eeq

With a gap in the boson spectrum, Eq.~(\ref{feqns}) imply that we can ignore the boson correlator in the determination of the 
fermion spectrum at small $\omega$. Indeed, the fermionic component of the equations are the same as those in Ref.~\cite{SY92}, 
and so we have the same gapless solution {\em i.e.\/}
\beq
G(\tau) = -\mbox{sgn} (\tau) \frac{A}{J^{1/2}} \left( \frac{\pi T}{|\sin (\pi T \tau)|} \right)^{1/2} \quad , \quad A = \frac{1}{(4 \pi)^{1/4}} \label{gappedferm}
\eeq
From Eq.~(\ref{beqns}) we can then obtain the long time behavior of the boson self energy
\beq
P (\tau) =  \frac{2 A^2 B \tilde{t}^2}{J} \frac{1}{(J |\tau|)^{1 + \gamma}}   e^{-m |\tau|} \quad, \quad |\tau| \gg 1/m ,~ T=0\,.
\eeq
From the analysis in Appendix~\ref{app:spectral}, as for Eq.~(\ref{bc1}), we conclude that $P(z)$ has a branch cut in the complex frequency plane at $z=\pm m$ with singular part
\beq
P_{\rm sing} (z) = \frac{2\pi A^2 B \tilde{t}^2}{J^2 \Gamma(1+\gamma)} \frac{i (z - m)^{\gamma}}{J^\gamma} \quad , \quad z \sim m \label{bc2}
\eeq
Comparing Eqs.~(\ref{bc1}) and (\ref{bc2}) with the Dyson equation in Eq.~(\ref{beqns}), it is not difficult to see that a consistent solution is only possible if
\beq
-m^2 + \chi_0^{-1} - P(z=m) + P(z=0) = 0
\eeq
and the exponent
\beq
\gamma = \frac{1}{2}.
\eeq
The dimensionless pre-factor $B$ is also determined to be
\beq
 (\tilde{t}/J^2)^2 A^2 B^2 = \frac{1}{4 \pi}\,.
 \eeq
 
\section{Composite operators}
\label{sec:fluc}

\subsection{$\lambda$ operator}
\label{sec:lambda}

Now we consider the structure of fluctuations
about the saddle point solutions described in the previous sections. First, we focus only on the fluctuations of the Lagrange multiplier field $\lambda_i$ about the saddle point value in 
Eq.~(\ref{lambda0}). This field represents the $\phi^2$ operator \cite{SSHiggs}, and so its scaling properties are important in determining the manner in which the gap 
in the $\phi$ spectrum opens up \cite{SSHiggs}, as we will discuss in Section~\ref{sec:num}.

We write
\beq
i\lambda_i = \chi_0^{-1} + P(i \omega_n = 0) + i \overline{\lambda}_i\,, 
\eeq
and then determine the effective action for $\overline{\lambda}_i$
fluctuations to leading order in large $N$ and large $M$, after integrating out the $f$ and the $\phi$ fields. 
The diagrams that contribute to this effective action are discussed in Appendix~\ref{app:lambda}, and they lead to an action of the form
\beq
\mathcal{S}_\lambda = \frac{T M'}{2} \sum_{\omega_n,i,j}\overline{\lambda}_i ( \omega_n) \left(
 \Pi_0^{ij} (\omega_n)  \,
 + \Pi_1^{ij} (\omega_n) \,
 \right)\overline{\lambda}_j (- \omega_n)  \,. \label{Slambda}
\eeq
Where we denote the bubble diagrams by $\Pi_0^{ij}$, which is diagonal in site index (i.e. $\Pi_0^{ij}=\Pi_0 \delta^{ij}$) and yields (in time domain):
\beq
\Pi_0 (\tau) =
\begin{tikzpicture}[scale=0.7, baseline={([yshift=-4pt]current bounding box.center)}]
\filldraw (-60pt,0pt) circle (1pt) node[left]{$i$};
\filldraw (60pt,0pt) circle (1pt) node[right]{$i$};
\draw[densely dashed] (-60pt,0pt) ..  controls (-15pt,15pt) and (15pt,15pt).. (60pt,0pt);
\draw[densely dashed] (-60pt,0pt) ..  controls (-15pt,-15pt) and (15pt,-15pt).. (60pt,0pt);
\end{tikzpicture}
=
\left[ \chi (\tau) \right]^2\,,
\eeq
where $\chi (\tau)$ is given by Eq.~(\ref{chicrit}).
We use $\Pi_1^{ij}$ to represent ladder diagrams with external indices $ij$. In general, we expect the matrix $\Pi_1^{ij}$ has permutation symmetry of the indices, which constrains the form of $\Pi_1^{ij}$ to be a matrix with identical diagonal elements and identical off-diagonal elements, i.e. there are only two free parameters. Such matrix admits one eigenvector that is uniform in site index with eigenvalue $(\Pi_1^{ii}+(N-1)\Pi_1^{ij})$ and $(N-1)$ non-uniform eigenvectors with eigenvalue $(\Pi_1^{ii}-\Pi_1^{ij})$. 
We are interested in the site-uniform mode, whose eigenvalue for the whole kernel including the bubble term can be written in the following symmetric way:
\begin{align}
    \Pi_0^{ii}+\Pi_1^{ii}+(N-1)\Pi_1^{ij} =\Pi_0 +  \frac{1}{N} \sum_{ij} \Pi_1^{ij} 
\end{align}
We denote the second term by $\Pi_1:=\frac{1}{N}\sum_{ij} \Pi_1^{ij} $ and computed in Appendix~\ref{app:lambda}, which requires evaluation of multiple infinite series of diagrams; they yield the result in Eq.~(\ref{eq:Pi1}):
\begin{align}
\Pi_1 (\tau) \approx
\frac{1}{2} \left(
    1-\frac{2\pi}{(\log \Lambda |\tau|)^2}
    \right)  \Pi_0(\tau)
 \,. 
\label{eq:Pi1m}
\end{align}
which is proportional to $\Pi_0(\tau)$ with a small $(\log \Lambda |\tau|)^{-2}$ correction.
Therefore we have the correlator for the
site-uniform $\bar{\lambda}$ fluctuation
\beq
\left\langle \overline{\lambda} (\omega_n)  \overline{\lambda} (-\omega_n) \right\rangle = \frac{1}{M'(\Pi_0 (\omega_n) + \Pi_1 ( \omega_n))} \,. 
\eeq
Limiting ourselves to the $\Delta_b=1/4$ critical state, to leading log accuracy at low frequency, this propagator is dominated by the Fourier transform of $\Pi_0 (\tau) \sim 1/|\tau|$, which yields
\beq
\left\langle \overline{\lambda} (\tau)  \overline{\lambda} (0) \right\rangle \sim \int d \omega \frac{e^{i \omega \tau}}{\ln( \Lambda/|\omega|)} \sim \frac{1}{|\tau|^{1 + \epsilon}}\,, 
\eeq
with $\epsilon = 1/(\ln (\Lambda |\tau|))$. So we can write the scaling dimension $[\overline{\lambda}] = (1+\epsilon)/2$, with $\epsilon$ representing logarithmic corrections to scaling.

\subsection{Electron operator} 

From the definition of the electron operator in Eq.~(\ref{defc}), we have to leading order in $1/N$ for the electron Green's function, $G_c$,
\bea
G_c (\tau) &=& G(\tau) \chi(\tau) \nn \\
&=& - \mbox{sgn}(\tau) \frac{FC}{J^2 |\tau|} \label{Gc}
\eea
where we have used Eq.~(\ref{chicrit}) and the exponent relation in Eq.~(\ref{Deltafb}).
Note that Eq.~(\ref{Deltafb}), and hence Eq.~(\ref{Gc}), hold for the both the gapless solutions in Sections~\ref{sec:deltag14} and~\ref{sec:delta14}. As was the case for the $\lambda$ fluctuations discussed above and in Appendix~\ref{app:lambda}, additional contributions to Eq.~(\ref{Gc}) from ladder diagrams only yield off-site terms which are
suppressed by $1/N$. So Eq.~(\ref{Gc}) is exact to leading order in the large $N$ limit of this paper.

It is remarkable that $G_c (\tau)$ has the same form as that in a Fermi liquid state. This can be seen to be a consequence of the relevance of the hopping, $t$, which moves single electrons between sites. However, it is important to note that despite the Fermi liquid form in Eq.~(\ref{Gc}), the states under considerations are not Fermi liquids: their elementary excitations are the fractionalized $f$ and $\phi$ excitations, which carry anomalous exponents.

\section{Numerical Results}
\label{sec:num}

We now present numerical tests of the solutions of Eqs.~(\ref{feqns})
and (\ref{beqns}). These go beyond the low frequency analytical analyses of Sections~\ref{sec:gapless} and~\ref{sec:gapped}, and include all frequencies. There are no ultraviolet divergencies, and so the solutions depend only upon the parameters in the Lagrangian.

Our numerical strategy was to pick at  first the values of the parameters $\widetilde{t}$ and $J$, and then make a choice for the boson susceptibility at zero frequency, $\chi_0$.
Then we iterate Eqs.~(\ref{feqns}) and (\ref{beqns}) until the solution converges. Finally, we insert the solution in Eq.~(\ref{gval}) and determine the value of $g$. So we determine $g$
as a function of $\chi_0$, rather than the other way around. 

\begin{figure}[h]
\begin{tabular}{c}
\includegraphics[width=3.75in]{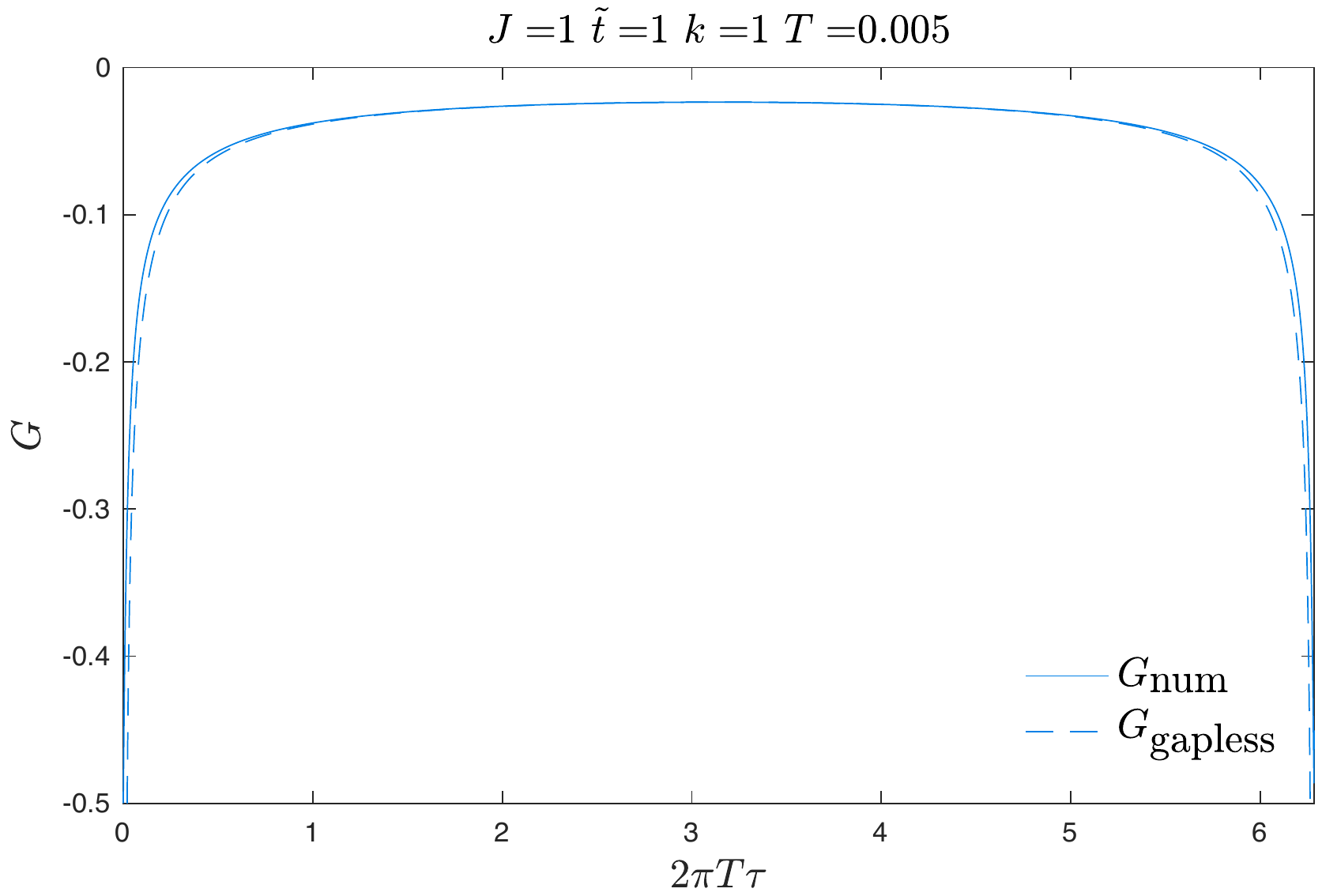}\\
\includegraphics[width=3.75in]{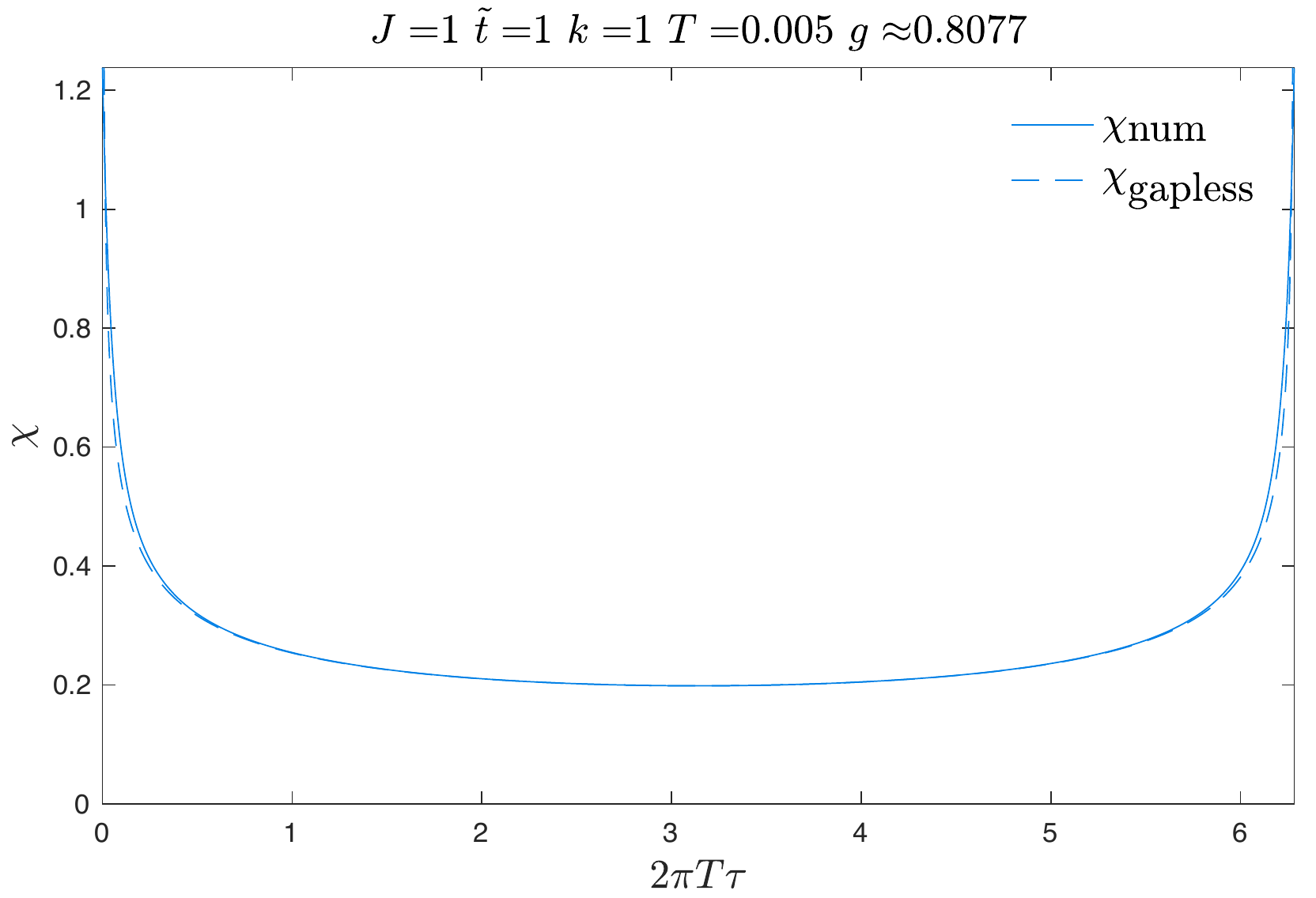}
\end{tabular}
\caption{The numerical results for $G(\tau)$ and $\chi(\tau)$ at the gapless phase are shown in solid lines for $J=1$, $\tilde{t}=1$, $k=1$, $T=0.005$ and $g\approx 0.8077$ (the input $\chi_0\approx 53.6$). The gapless conformal answers for $G$ and $\chi$ are plotted  in dashed lines with the values from Eq.~(\ref{alphaexp}), $\Delta_f = 1/3$, $\Delta_b = 1/6$. The prefactor $C$ is determined by (\ref{chi0c}) with $\Delta_b=1/6$, then $F$ is determined by (\ref{eq:FC}).}
\label{fig:GandChigapless}
\end{figure}
First, we examined the gapless solutions, with the input $\chi_0$, the conformal solution prefactor is determined by (\ref{chi0c}). 
A solution with $\Delta_f > 1/4$ is shown in Fig.~\ref{fig:GandChigapless}.
In this case, at any finite temperature, although the prefactor equations (\ref{eq:FC}) from the saddle point equations do not determine the prefactors $F$ and $C$ separately, but the matching condition (\ref{chi0c}) determines them. We can think about it this way: different $\chi_0$ results from different $g$ and it determines different $F$ and $C$. So such a gapless solution can be obtained for a range of values of $g$ at a fixed $\Delta_f$. And at zero temperature, when $\chi_0$ diverges, we cannot determine $F$ and $C$ separately. Thus this gapless solution defines a critical phase.

\begin{figure}[h]
\begin{tabular}{c}
\includegraphics[width=3.75in]{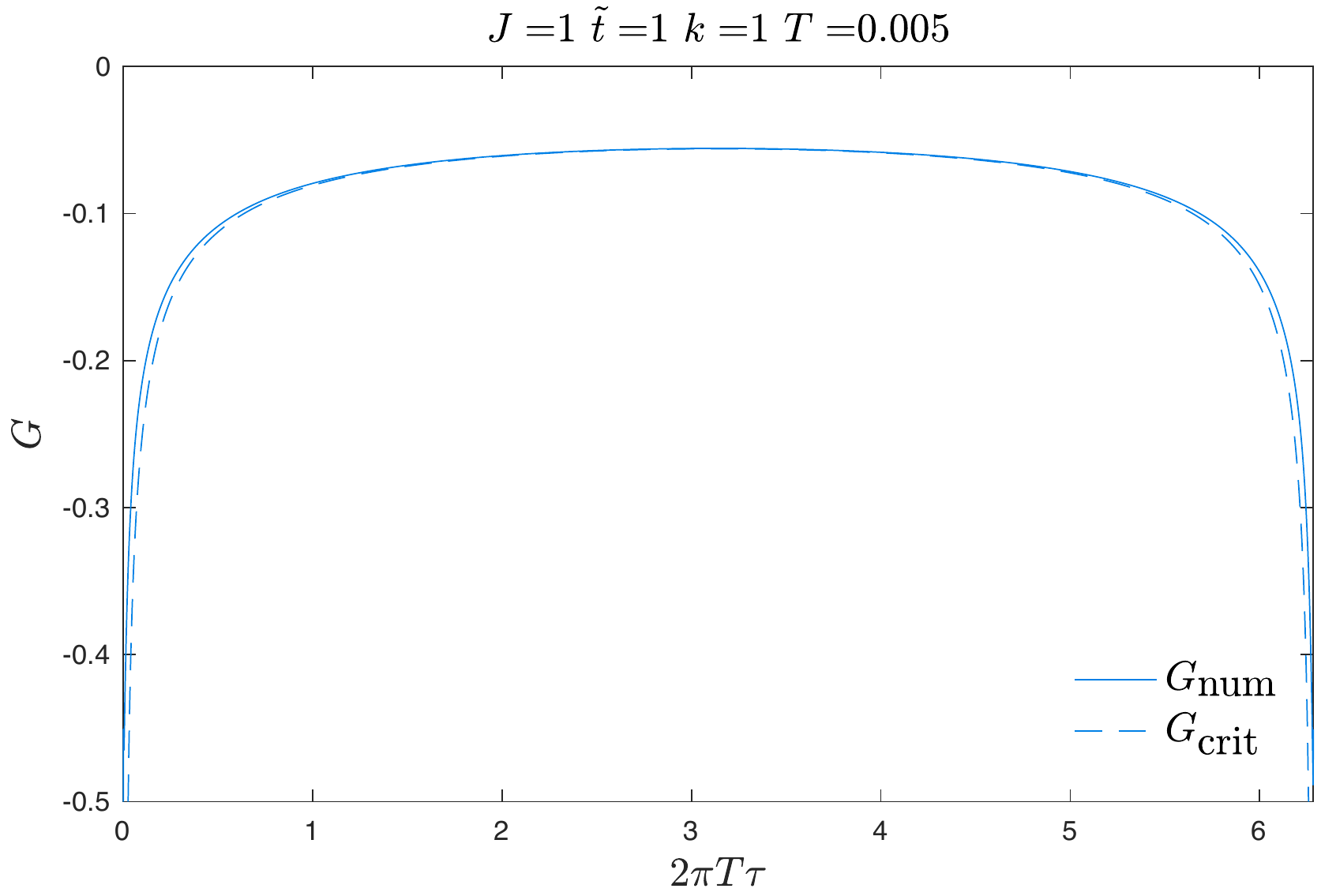}\\
\includegraphics[width=3.75in]{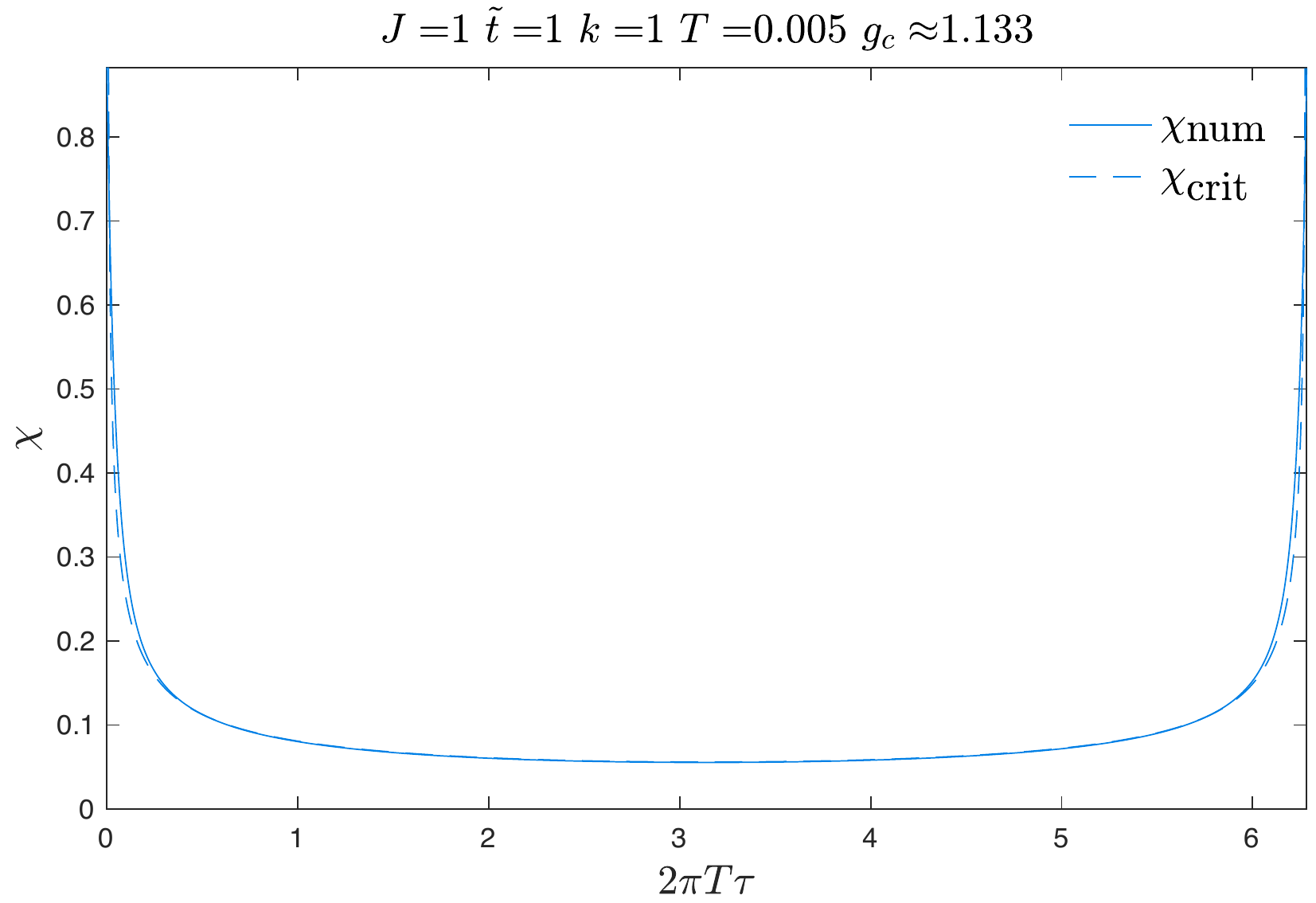}
\end{tabular}
\caption{The numerical results for $G(\tau)$ and $\chi(\tau)$ at the critical phase are shown in solid lines for $J=1$, $\tilde{t}=1$, $k=1$, $T=0.005$ and $g_{c}\approx 1.133$ (The input $\chi_0\approx 18.6$ can be obtained from (\ref{chi0c}) with $\Delta_b=1/4$). The critical conformal answers for $G$ and $\chi$ are plotted  in dashed lines.}
\label{fig:GandChicrit}
\end{figure}
Next, we examined the gapless solution with $\Delta_f=1/4$ in Fig.~\ref{fig:GandChicrit}. 
In this case, the saddle point equations determine $F$ and $C$ separately in the prefactor equations Eq.~(\ref{e2}). For each value of $\widetilde{t}$, $J$, $k$ and $T$, the critical susceptibility is determined by (\ref{chi0c}), thus it determines an unique $g_c$.

We also examined the $T$ dependence of $\chi_0^{-1}$ predicted by Eq.~(\ref{chi0c}). We choose different values of $T$ with other parameters fixed, and found a $T$-indpendent value of $g$. This confirms the analysis in Section~\ref{sec:gaplessT} on extending the $T=0$ gapless solution to nonzero $T$.

\begin{figure}[h]
\begin{tabular}{c}
\includegraphics[width=4.45in]{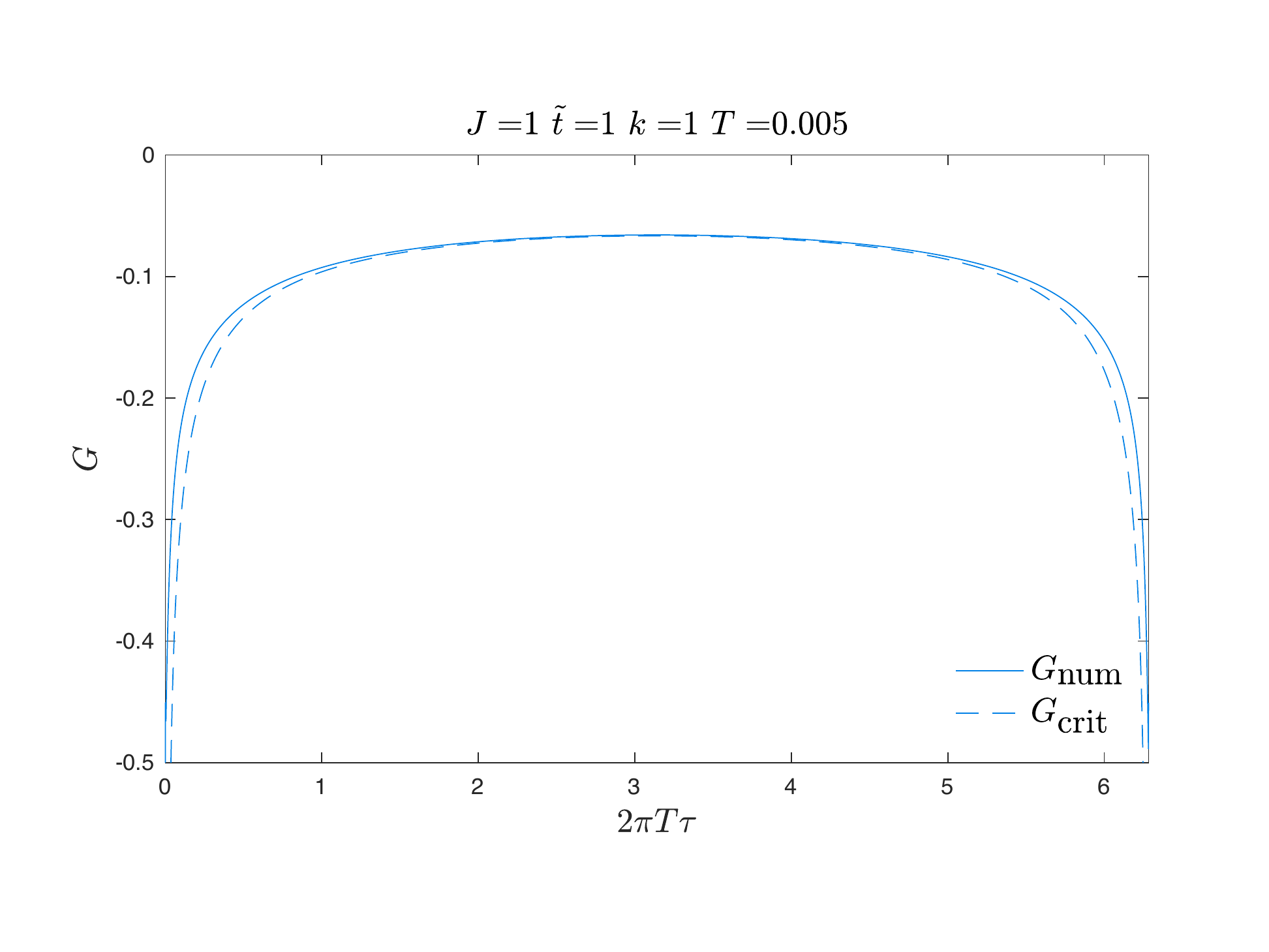}\\
\includegraphics[width=3.75in]{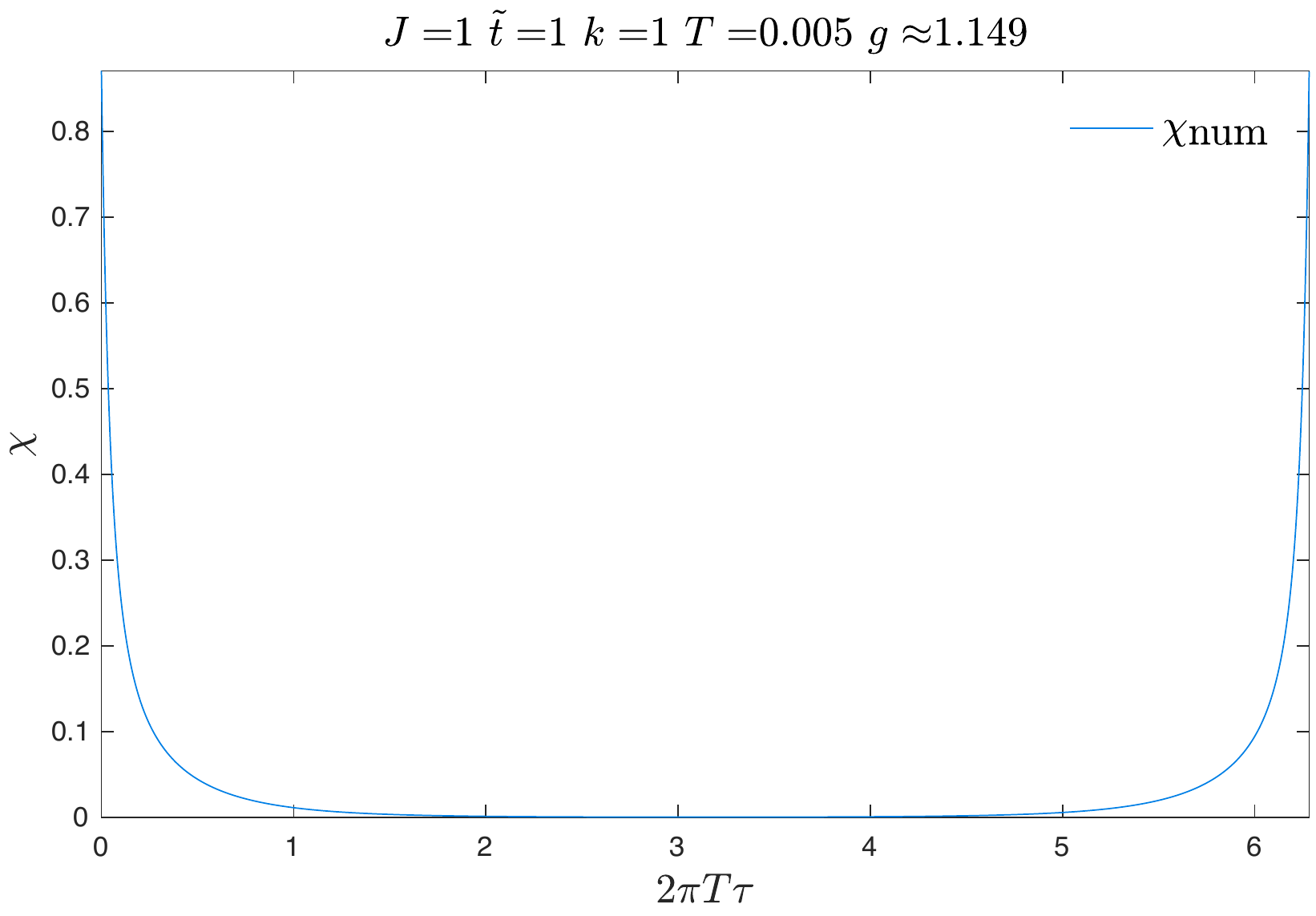}
\end{tabular}
\caption{The numerical results for $G(\tau)$ and $\chi(\tau)$ at the gapped boson phase are shown in solid lines for $J=1$, $\tilde{t}=1$, $k=1$, $T=0.005$ and $g\approx 1.149$ (The input $\chi_0\approx 6.6$). The critical conformal answers for $G$ from (\ref{gappedferm}) is plotted  in dashed line.}
\label{fig:GandChigapped}
\end{figure}
Finally, we examined the gapped boson solution of Section~\ref{sec:gapped} in Fig.~\ref{fig:GandChigapped}. The normalization constant for the fermion conformal answer is defined in Eq. (\ref{gappedferm}).
Again we find good agreement between the numerical solution and our analytic form.

\begin{figure}[h]
\begin{center}
\includegraphics[width=4.5in]{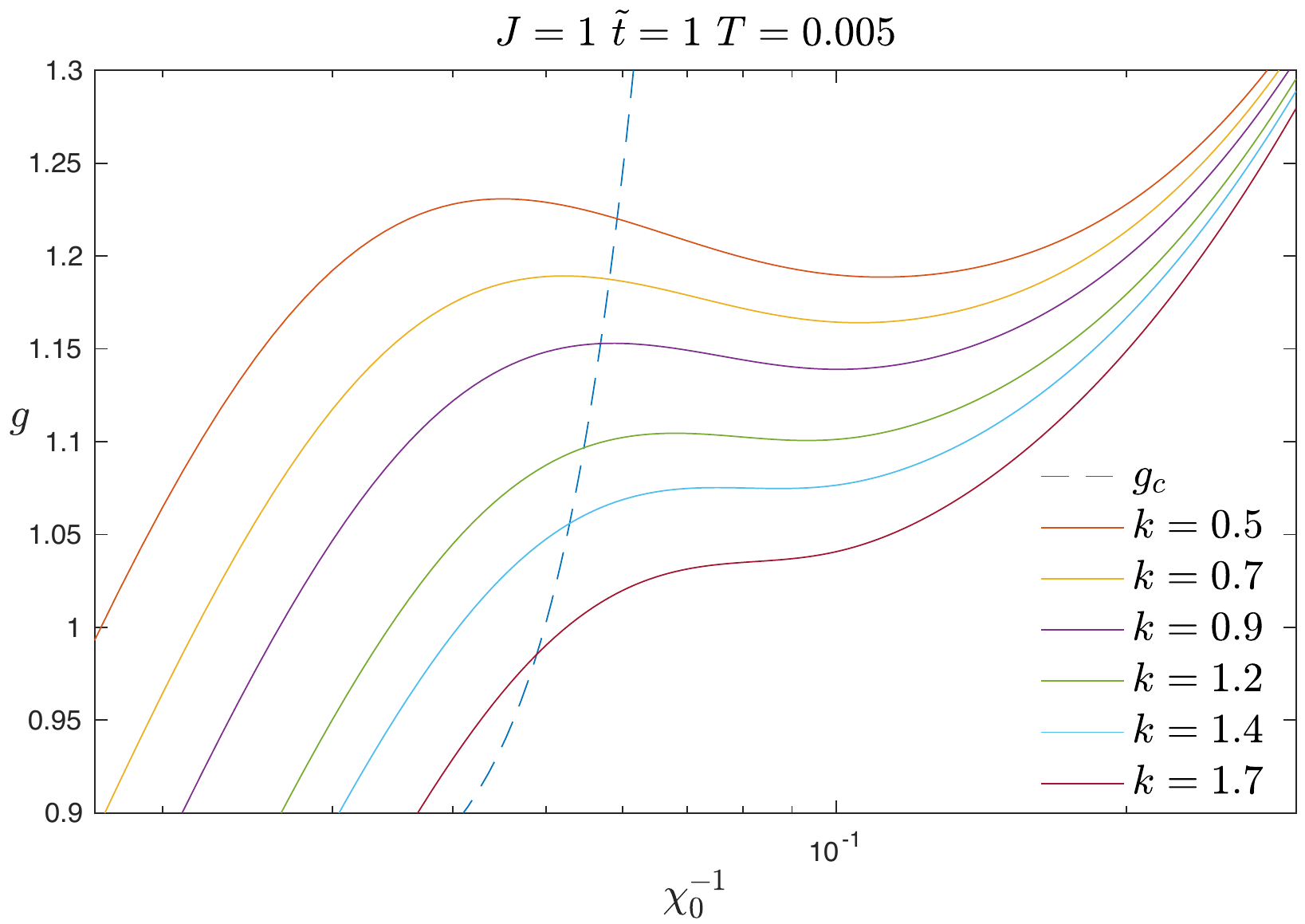}
\end{center}
\caption{The numerical results for $g$ as function of $\chi_{0}^{-1}$ for various values of the parameter $k$ at $J=1$, $\tilde{t}=1$ and $T=0.005$. The critical $g_{c}$ is shown in dashed line. For $k<k_{c}\approx 0.9$ the critical point is located inside an unphysical domain, which means that  the critical phase is absent and there is first-order phase transition.  }
\label{fig:gofk}
\end{figure}
Now we turn to determining how the various solutions fit together in a phase diagram as a function of $k$ and $g$. We set $\widetilde{t}=J=1$
in this analysis. As our independent parameter is $\chi_0$, and not $g$, we show in Fig.~\ref{fig:gofk} the value of $g$ determined from Eq.~(\ref{gval}) as a function of $\chi_0^{-1}$ for various values of $k$. The values of $\chi_0^{-1}$ corresponding to Eq.~(\ref{chi0c}) at $\Delta_b=1/4$ yield the value of $g_c$ for each $k$: this is plotted as the dashed line. The most notable feature of Fig.~\ref{fig:gofk} is the non-monotonic dependence of $g$ on $\chi_0^{-1}$ for certain $k$. This implies that for a given $g$ there are multiple solutions of the saddle point equations
in Eqs.~(\ref{feqns}), (\ref{beqns}), and (\ref{gval}) corresponding to the different solutions for the value of $\chi_0$. To distinguish between the solutions, we have to evaluate the free energy of each solution and pick the one with the lowest free energy. We have not carried out this evaluation, 
and so are unable to determine the precise location of the transition between the gapless and gapped solutions. In any case, we can conclude that there is a first-order transition from the gapless to the gapped solution when $g$ is a decreasing function of $\chi_0^{-1}$ near $g_c$.

\begin{figure}[h!]
\begin{tabular}{cc}
\includegraphics[width=3.5in]{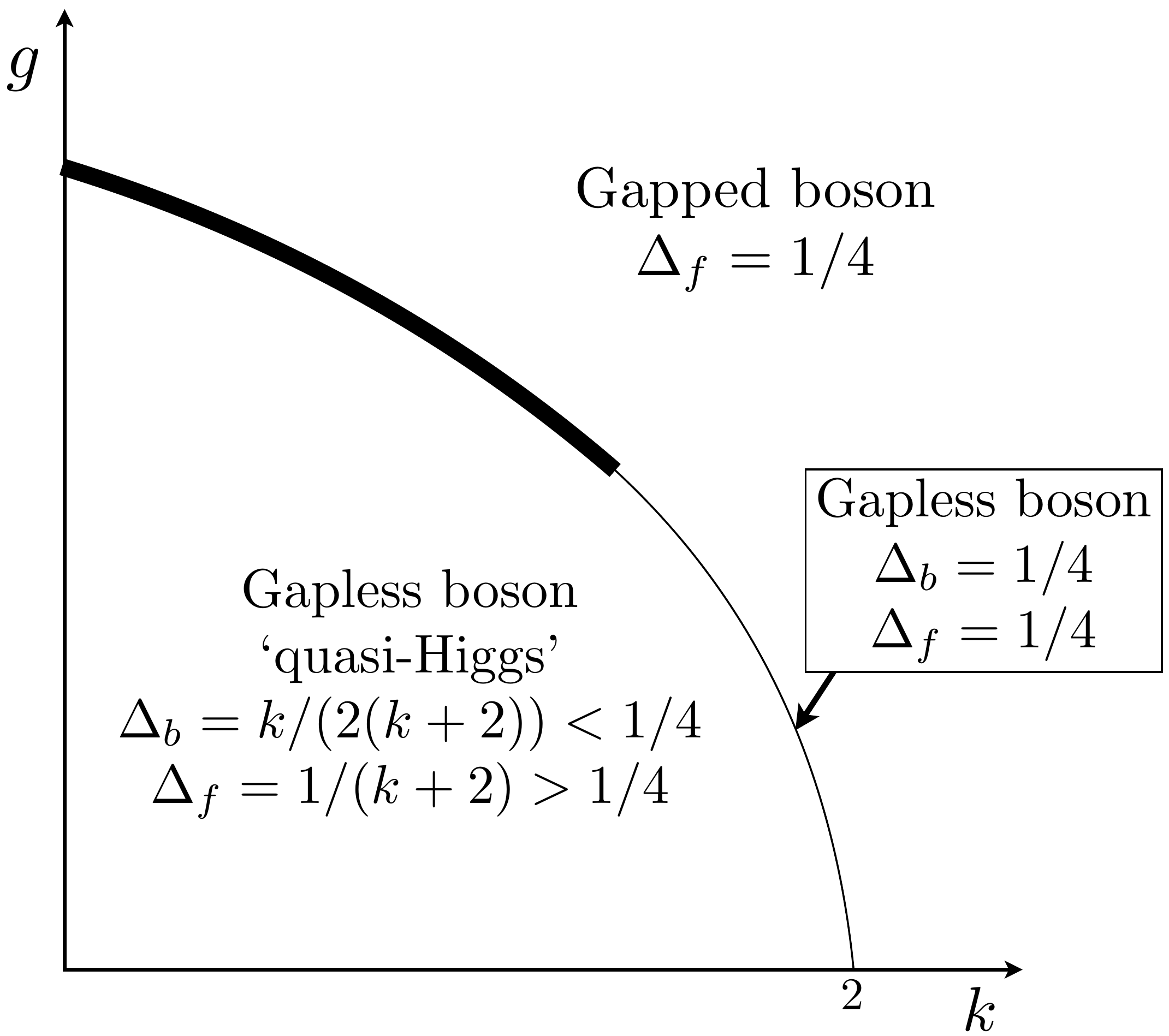}
\end{tabular}
\caption{Schematic phase diagram. The fermions are gapless in all phases. 
The thick line indicates a first-order transition, while the thin line is a $\Delta_b=\Delta_f=1/4$ critical state.}
\label{fig:phasediag}
\end{figure}

On the basis of the above analysis and Fig.~\ref{fig:gofk}, we assemble the schematic phase diagram in Fig.~\ref{fig:phasediag}. The gapless phase with $\Delta_f > 1/4$ is separated from the gapped boson phase by either a first-order or a second-order phase transition. For the latter case, the critical state is described by the $\Delta_f=1/4$ solution described in Section~\ref{sec:delta14}.

\subsection{Small gap scaling}
\label{sec:scaling}

We now examine the nature of the scaling properties of the gapped side of the second-order
transition in Fig.~\ref{fig:phasediag}. On general grounds, we introduce the exponent
$z$ by assuming that the boson energy gap, $m$, vanishes as 
\beq
m \sim (g-g_c)^z \quad , \quad T=0\,,\label{s4}
\eeq
As the energy gap appears from a $\sim \phi^2$ perturbation of the critical theory, we expect that the scaling dimension of $\phi^2$ is related to $z$ via
$[\phi^2] = 1 - 1/z$.
On the other hand, as explained in the context of the Wilson-Fisher CFT in Ref.~\cite{SSHiggs},
in the large $M$ expansion $[\lambda] = [\phi^2]$ and so
\beq
[\lambda] = 1 - \frac{1}{z}
\label{lambdaz}
\eeq
We examined the scaling dimension of $\lambda$ in Section~\ref{sec:lambda}, and found that
$[\overline{\lambda}] = (1+\epsilon)/2$, with $\epsilon$ representing logarithmic corrections to scaling. So $z = 2(1+ \epsilon)$.

From our numerical solutions, it turned to be difficult to obtain accurate values of the boson gap, $m$, to test the above scaling predictions. So we employed an alternative method, which examined the full functional form of the boson susceptibility $\chi (\tau)$.
From the structure of the gapped solution in Eq.~(\ref{bc3}) we can expect a scaling solution for the $T=0$ susceptibility of the form
\beq
\chi (\tau) = \left(\frac{m}{J^3}\right)^{1/2} \Phi_1 (m \tau) \label{bc4}
\eeq
for some scaling function $\Phi_1$. Clearly, Eq.~(\ref{bc4}) is compatible with the long-time limit in Eq.~(\ref{bc3}).
Then integrating Eq.~(\ref{bc4}) over $\tau$, we obtain the divergence of the static susceptibility as the gap, $m$, vanishes
\beq
\chi_0 \sim m^{-1/2}\quad , \quad T=0\,.
\label{s1}
\eeq
For a second-order transition to a gapless phase, with the critical point described by the $\Delta_b = 1/4$ solution in Section~\ref{sec:delta14},
Eq.~(\ref{chi0c}) implies that the static susceptibility behaves as
\beq
\chi_0 \sim T^{-1/2} \quad, \quad m \rightarrow 0\,.
\label{s2}
\eeq
Combining Eqs.~(\ref{s1}) and (\ref{s2}), we propose the scaling form
\beq
\chi_0 = T^{-1/2} \Phi_2 (m/T)\,. \label{s3}
\eeq
In our numerical solution, Eq.~(\ref{s3}) is difficult to test directly because we treat $\chi_0$ as an independent parameter and compute $m$ and $g$,
and also $m$ is only defined at $T=0$. 
As we can also measure the deviation from criticality by $\chi_0^{-1}$, we can combine the scaling in Eqs.~(\ref{s4}) and (\ref{s3}) to write
\beq
g - g_c = T^{1/z} \Phi_3 (\chi_0 T^{1/2}) \quad, \quad g>g_c \,, \label{s5}
\eeq
where $\Phi_3$ is another scaling function. 
Eq.~(\ref{s5}) is now expressed in a form which is adapted to our numerical approach: we pick the values of $\chi_0$ and $T$, and compute $g$.
Also, we can compute the value $g_c$ by requiring that Eq.~(\ref{s5}) be compatible with Eq.~(\ref{chi0c}) {\em i.e.\/} 
\beq
\Phi_3 (x) = 0~\mbox{at}~x=\frac{C  \widetilde{\Pi} (1/2) \Gamma ( 1/4 )}{J^{3/2} \Gamma  ( 3/4 ) }
\eeq

\begin{figure}[h]
\begin{tabular}{c}
\includegraphics[width=4.20in]{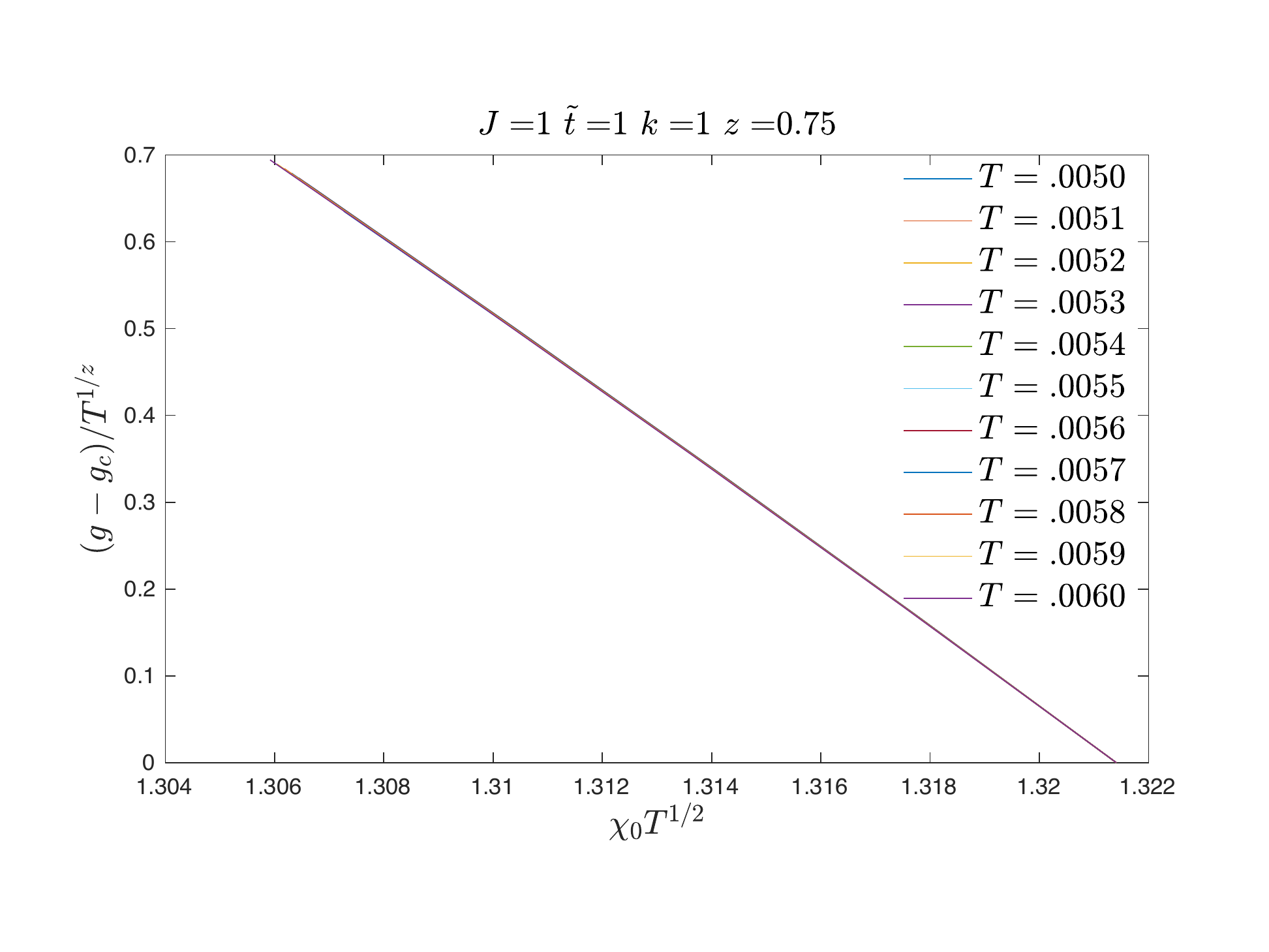}\\
\includegraphics[width=3.75in]{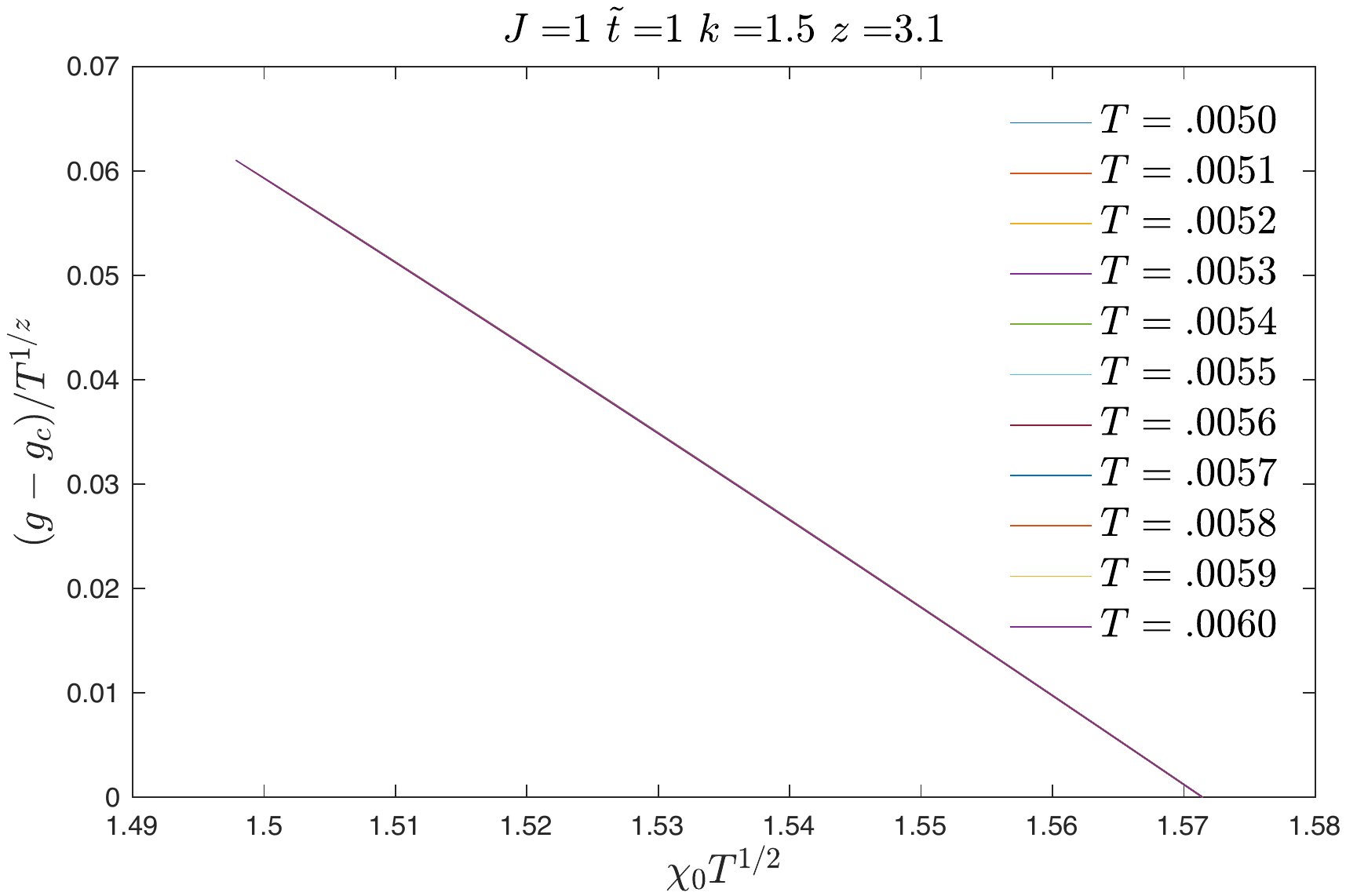}\\
\includegraphics[width=3.75in]{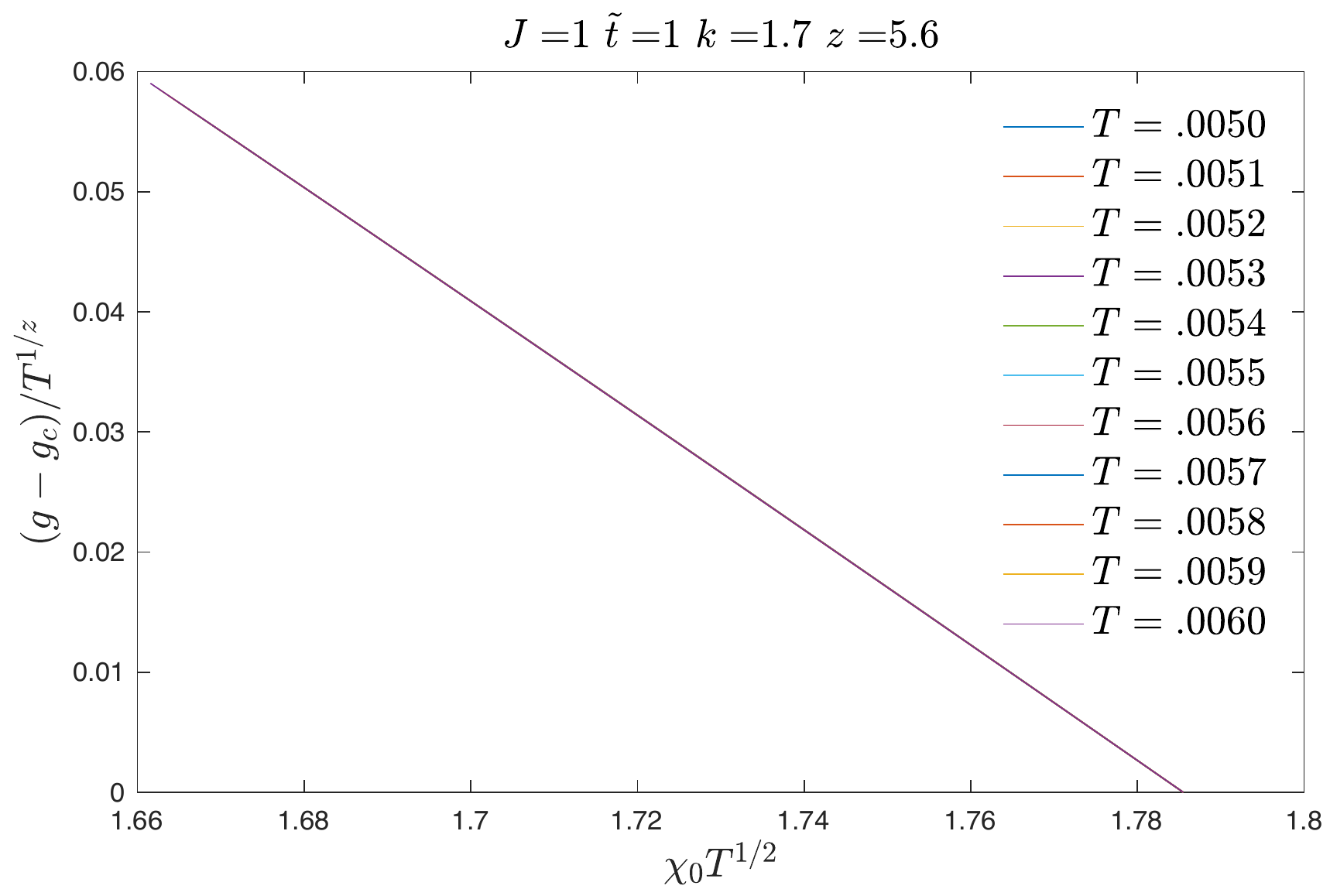}
\end{tabular}
\caption{Tests of the scaling in Eq.~(\ref{s5}): numerical plots for $(g-g_{c})/T^{1/z}$ as a function of $\chi_{0}T^{1/2}$ at $J=1$, $\tilde{t}=1$ and $k=1$, $k=1.5$ and  $k=1.7$. We see that at $z=0.75$, $z=3.1$ and $z=5.6$ all lines almost exactly overlap, as expected in the critical region.  }
\label{fig:Scaleplots}
\end{figure}
\begin{figure}[h]
\begin{tabular}{c}
\includegraphics[width=4.20in]{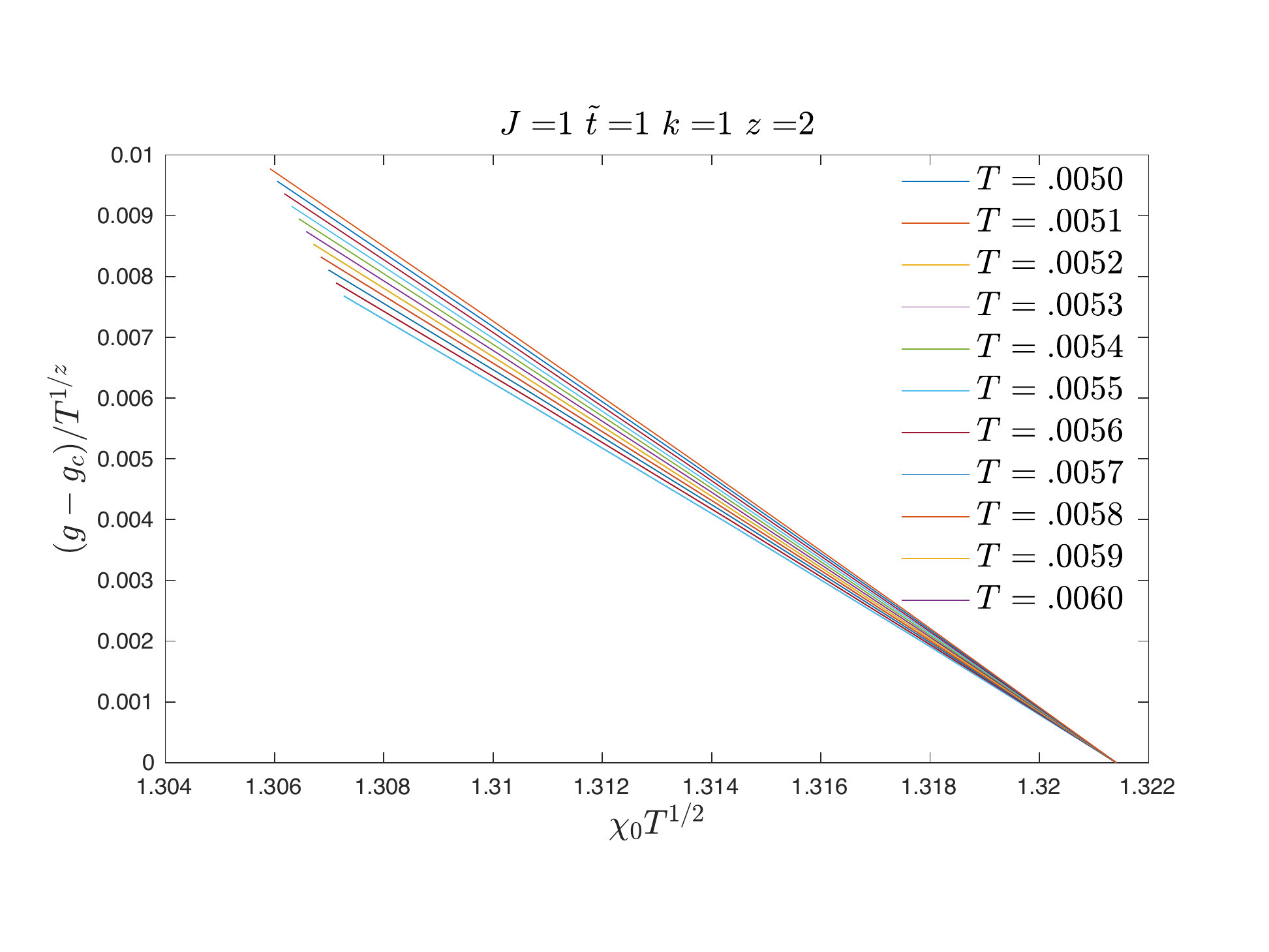}\\
\includegraphics[width=3.75in]{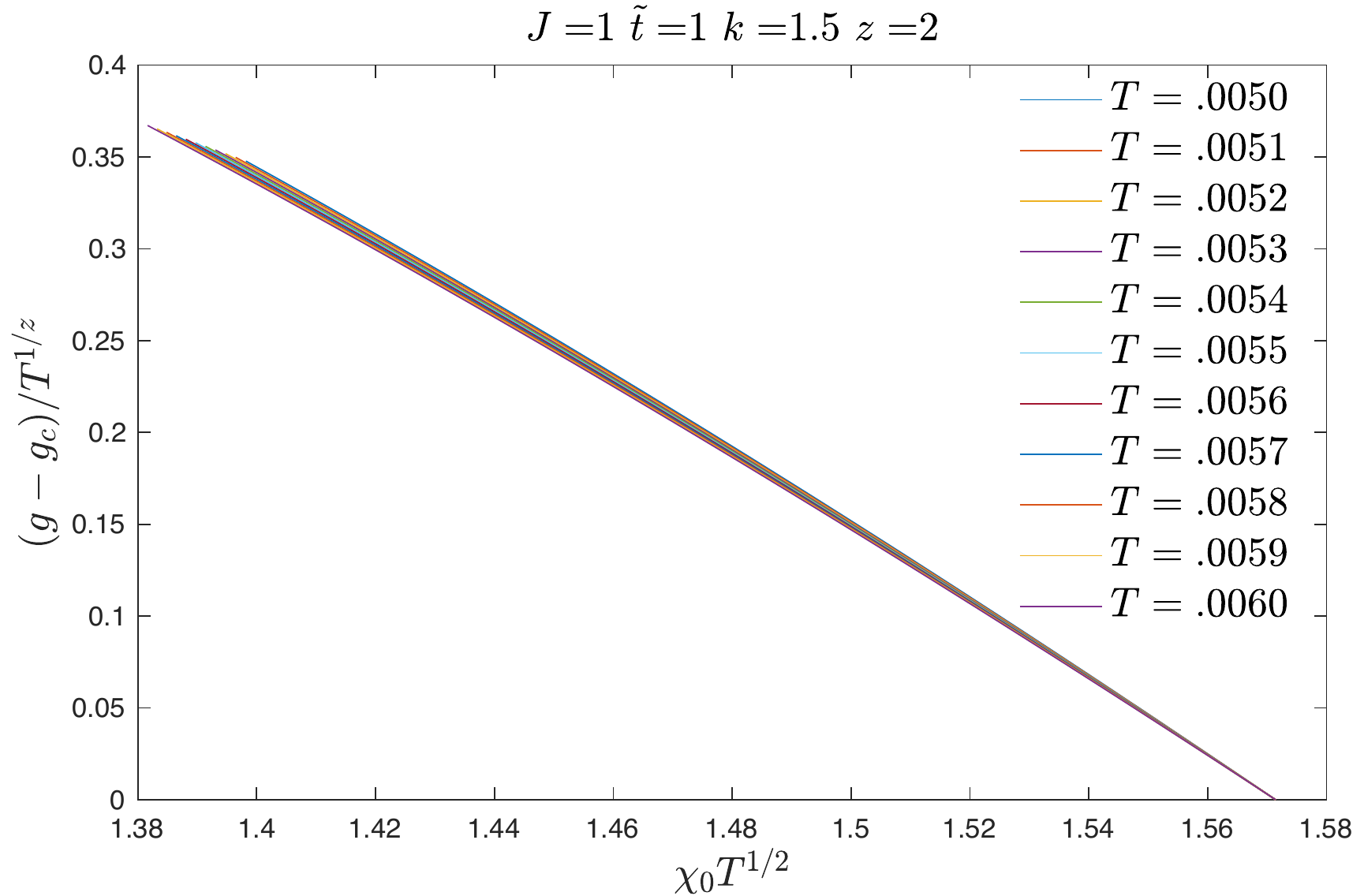}\\
\includegraphics[width=3.75in]{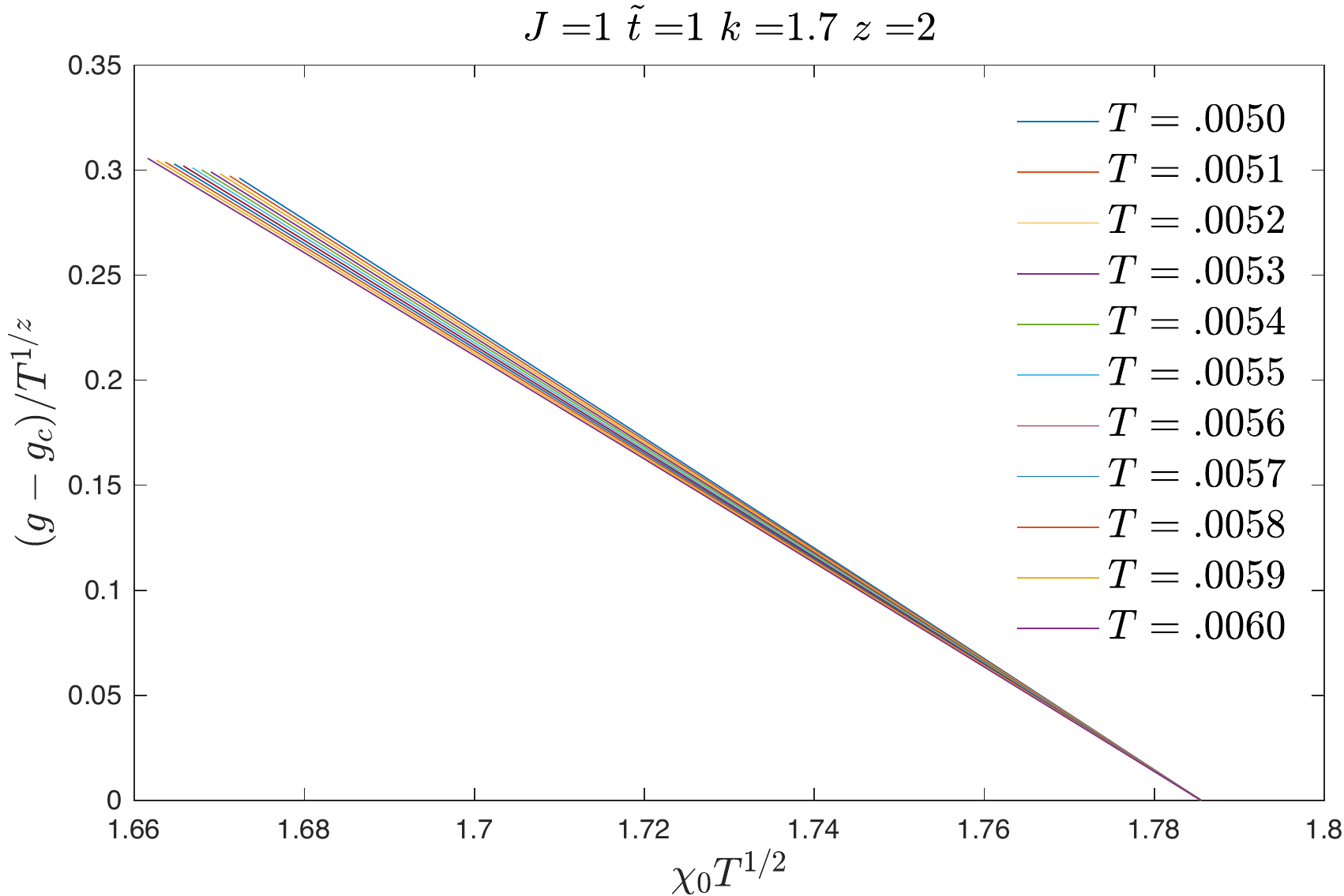}
\end{tabular}
\caption{The same plots as in Fig.~\ref{fig:Scaleplots}, but now scaled with $z=2$.}
\label{fig:Scaleplotsz2}
\end{figure}
We show tests of the scaling in Eq.~(\ref{s5}) in Figs.~\ref{fig:Scaleplots} and~\ref{fig:Scaleplotsz2}.
We find that scaling as a function of $\chi_0 T^{1/2}$ is extremely well obeyed, confirming that the critical state is described by $\Delta_f=\Delta_b = 1/4$\footnote{We have also examined the first order transition region and would not find a scaling behavior as expected}. 
Specifically, we verified that at $g=g_c$, the right hand side of Eq.~(\ref{s5}) was $T$ independent as $T$ was varied while keeping $\chi_0 T^{1/2}$ fixed.

On the other hand, scaling with $(g-g_c)/T^{1/z}$ yields variable values of $z$ depending upon the value of $k$, and of the window of parameters used for the scaling plots, as is apparent from Fig.~\ref{fig:Scaleplots}. We generally obtained values of $z>2$, except at values of $k$ near the onset of the first order transition. Fig.~\ref{fig:Scaleplotsz2} shows that scaling with $z=2$ yields reasonable data collapse, with deviations which appear to be within the range of 
logarithmic corrections described in Section~\ref{sec:lambda}. However, data
collapse could be improved with larger values of $z$ especially by focusing on values of $g$ very close to $g_c$: it does not appear these large and variable
values of $z$ are meaningful. More precise tests of the nature of the phase transitions requires a detailed knowledge of the structure of the logarithmic corrections, which we have not computed here.

\section{Conclusions}
\label{sec:conc}

Our paper has presented an exactly solvable model of fractionalization in metallic states in the presence of disorder and interactions. We considered a $t$-$J$ model in which
the electron, $c$, fractionalizes into a fermion $f$ and a boson $\phi$ both carrying
$\mathbb{Z}_2$ gauge charges. As the fermions $f$ carries both the global U(1) charge and SU(2) spin of the electron, these fractionalized phases can be considered as realizations of the
`orthogonal metal' of Ref.~\cite{OM12}.

The phase diagram of our model is schematically presented in Fig.~\ref{fig:phasediag}. 
There are two extended phases, separated either by a first order transition, or a critical line.

In one phase, the boson $\phi$ is gapped, while the fermion $f$ is gapless. This implies that
the electron $c \sim f \phi$ also has a gapped spectral function. On the other hand, thermodynamic properties are largely controlled by the gapless  fermions. We propose this gapped boson phase as a toy model for the pseudogap regime of the cuprates.

In the other phase, and also on the critical line, both the fermions and bosons are gapless, and decay with time as $|\tau|^{-2 \Delta_{f,b}}$, where the values of the exponents are specified
in Fig.~\ref{fig:phasediag}. In a Higgs phase, in which the $\mathbb{Z}_2$ charges are confined, the boson correlator would decay to a non-zero constant. As the boson decay here is a power-law in time, we labeled this phase as a quasi-Higgs phase. 

One of the most interesting properties of the quasi-Higgs phase follows from the exponent identity in Eq.~(\ref{Deltafb}). The Green's function of the electron operator, $c$, decays
with time as $1/\tau$ (Eq.~(\ref{Gc})), which is the form of the local Green's function in a Fermi liquid. This result is a consequence of the relevance of the hopping term, $t$, in the Hamiltonian which transfers single electrons between sites. 
Unlike previous SYK models, the present model balances the hopping ($t$) and interaction ($J$) terms against each other, rather than one of them dominating; this leads 
to the Fermi liquid form of the one-electron Green's function. However, despite this form, most other properties are not Fermi liquid-like {\em e.g.\/}
the spin susceptibility is dominated by the response of the $f$ fermions which have an
anomalous scaling dimension $\Delta_f$. These intruiging properties are suggestive of the overdoped regime of the cuprates, where there are indications of an extended non-Fermi liquid regime, although photoemission indicates a well-formed large Fermi surface \cite{Damascelli05,Hussey09,Bozovic18,Armitage18}.

Extending our toy model to a more realistic model of the cuprates requires introducing spatial structure and examining transport properties. A number of methods of doing so have been introducing recently \cite{Gu2017local,Davison17,Gu17,Balents17,Patel18,DC18} for the SYK model, and it would be interesting to apply these, or others, to models similar to the one presented here.

It would also be useful to examine holographic duals of the phases and phase transitions presented here. Given the mapping of the SYK model to AdS$_2$ gravity \cite{SS10,nearlyads2,kitaev2017}, 
and the conformal invariance of the gapless solution in Section~\ref{sec:gaplessT}, it seems plausible that such holographic duals are possible. The AdS$_2$ phase transitions studied in Refs.~\cite{Iqbal10,Iqbal11} are likely candidates for developing the dual theory.

\subsection*{Acknowledgements}

We thank Hong Liu, Aavishkar Patel and Yi-Zhuang You for useful discussions.
This research was supported by the National Science Foundation under Grant No. DMR-1360789. 
Research at Perimeter Institute is supported by the Government of Canada through Industry Canada and by the Province of Ontario through the Ministry of Research and Innovation. SS also acknowledges support from Cenovus Energy at Perimeter Institute. 
YG is supported by the Gordon and Betty Moore Foundation EPiQS Initiative through Grant (GBMF-4306). GT is supported  by the MURI grant W911NF-14-1-0003 from ARO and by DOE grant de-sc0007870.

\appendix

\section{Spectral functions}
\label{app:spectral}

We recall a few basic facts about fermion and boson spectral functions. 

At the Matsubara frequencies, the fermion Green's function is defined by
\bea
G (\tau) &=& - \left\langle T \left( f(\tau) f^\dagger (0) \right) \right\rangle \nn \\
G(i  \omega_n) &=& \int_0^{1/T} d \tau e^{i \omega_n \tau} G(\tau),
\eea
and this is continued to all complex frequencies $z$ via
the spectral representation
\beq
G(z) = \int_{-\infty}^{\infty} \frac{d \Omega}{\pi} \frac{\rho (\Omega)}{z - \Omega}. \label{spec}
\eeq
The spectral density $\rho (\Omega ) > 0$ for all real $\Omega$ and $T$. The retarded Green's function is $G^R (\omega) = G(\omega + i \eta)$ with $\eta$ a positive infinitesimal, while the advanced Green's function is $G^A (\omega) = G(\omega - i \eta)$. From these expressions we obtain
\beq
G(\tau) = - \int_{-\infty}^{\infty} \frac{d \Omega}{\pi} \rho(\Omega) \, \frac{ e^{(\beta - \tau) \Omega}}{e^{\beta \Omega} + 1} \quad , \quad 0 < \tau < \beta \,. \label{Gtau}
\eeq
So in the limit $T \rightarrow 0$ we have
\beq
G(\tau) = \left\{ \begin{array}{ccc} \displaystyle - \int_0^{\infty} \frac{d\Omega}{\pi} \rho (\Omega) \, e^{- \Omega \tau} & ~,~& \tau > 0 \\
~\\
 \displaystyle  \int_0^{\infty} \frac{d\Omega}{\pi} \rho (-\Omega) \, e^{ \Omega \tau} & ~,~& \tau < 0 \,.
\end{array}
\right.
\eeq
We will focus on the particle-hole symmetric case, in which case $\rho (\Omega ) = \rho (-\Omega)$. 

For bosons, the Green's function is defined by
\bea
\chi (\tau) &=& \left\langle T \left( \phi (\tau) \phi (0) \right) \right\rangle \nn \\
\chi (i  \omega_n) &=& \int_0^{1/T} d \tau e^{i \omega_n \tau} \chi (\tau),
\eea
and we have the spectral representation
\beq
\chi (z) = \int_{-\infty}^{\infty} \frac{d \Omega}{\pi} \frac{ \zeta (\Omega)}{z - \Omega}. \label{specb}
\eeq
Now the positivity condition is $\Omega \zeta (\Omega) < 0$. For the real bosons $\zeta (-\Omega) = - \zeta (\Omega)$,
and we will assume this from now. The analog of Eq.~(\ref{Gtau}) is 
\beq
\chi (\tau) = - \int_{-\infty}^{\infty} \frac{d \Omega}{\pi} \zeta (\Omega) \, \frac{ e^{(\beta - \tau) \Omega}}{e^{\beta \Omega} - 1} \quad , \quad 0 < \tau < \beta \,. \label{Gtaub}
\eeq
So in the limit $T \rightarrow 0$ we have
\beq
\chi (\tau) = \left\{ \begin{array}{ccc} \displaystyle - \int_0^{\infty} \frac{d\Omega}{\pi} \zeta (\Omega) \, e^{- \Omega \tau} & ~,~& \tau > 0 \\
~\\
 \displaystyle  \int_0^{\infty} \frac{d\Omega}{\pi} \zeta (-\Omega) \, e^{ \Omega \tau} & ~,~& \tau < 0 \,.
\end{array}
\right.
\label{ab1}
\eeq

\section{Diagrammatic summation for site-uniform $\bar{\lambda}$ fluctuations}
\label{app:lambda}

In this appendix, we consider the diagrams for site-uniform $\bar{\lambda}$ fluctuation drawn in the time domain. It turns out the leading diagrams in large $N,M,M'$ are all horizontal ladders, which can be summed using a recurrence relation. 

\subsection{The index structure of diagrams}
The original Lagrangian relevant for the vertices before averaging is:
\begin{align}
\frac{1}{\sqrt{NM}} \left( 
\sum_{i,j,p,\alpha } t_{ij } \phi_{i,p} \phi_{j,p} f^\dagger_{i,\alpha }f_{j,\alpha} + \sum_{i>j , \alpha, \beta} J_{ij} f^\dagger_{i,\alpha} f_{i,\beta} f^\dagger_{j,\beta} f_{j,\alpha}
\right)
\end{align}
There are in general four types of ladders after averaging:
\begin{align}
\begin{tikzpicture}[scale=0.7, baseline={([yshift=-4pt]current bounding box.center)}]
\draw[densely dashed] (0pt,30pt) -- (60pt,30pt);
\draw[densely dashed](0pt,-30pt) -- (60pt,-30pt);
\filldraw[fill=black] (30pt,30pt) circle (1pt)node[above]{$t$} ;
\filldraw[fill=black] (30pt,-30pt) circle (1pt) node[below]{$t$};
\draw (30pt,30pt) .. controls (35pt,10pt) and (35pt,-10pt).. (30pt,-30pt);
\draw (30pt,30pt) .. controls (25pt,10pt) and (25pt,-10pt).. (30pt,-30pt);
\draw [->,>=stealth] (26.3pt,-1pt) -- (26.3pt,1pt);
\draw [-<,>=stealth] (33.7pt,-1pt) -- (33.7pt,1pt);
\end{tikzpicture}
\qquad 
\begin{tikzpicture}[scale=0.7, baseline={([yshift=-4pt]current bounding box.center)}]
\draw (0pt,30pt) -- (60pt,30pt);
\draw(0pt,-30pt) -- (60pt,-30pt);
\filldraw[fill=black] (30pt,30pt) circle (1pt)node[above]{$t$} ;
\filldraw[fill=black] (30pt,-30pt) circle (1pt) node[below]{$t$};
\draw[densely dashed] (30pt,30pt) .. controls (35pt,10pt) and (35pt,-10pt).. (30pt,-30pt);
\draw[densely dashed] (30pt,30pt) .. controls (25pt,10pt) and (25pt,-10pt).. (30pt,-30pt);
\draw [->,>=stealth] (15pt,30pt) -- (16pt,30pt);
\draw [-<,>=stealth] (15pt,-30pt) -- (16pt,-30pt);
\draw [->,>=stealth] (45pt,30pt) -- (46pt,30pt);
\draw [-<,>=stealth] (45pt,-30pt) -- (46pt,-30pt);
\end{tikzpicture}
\qquad 
\begin{tikzpicture}[scale=0.7, baseline={([yshift=-4pt]current bounding box.center)}]
\draw (0pt,30pt) -- (30pt,30pt);
\draw(0pt,-30pt) -- (30pt,-30pt);
\draw[densely dashed] (60pt,30pt) -- (30pt,30pt);
\draw[densely dashed] (60pt,-30pt) -- (30pt,-30pt);
\filldraw[fill=black] (30pt,30pt) circle (1pt)node[above]{$t$} ;
\filldraw[fill=black] (30pt,-30pt) circle (1pt) node[below]{$t$};
\draw (30pt,30pt) .. controls (35pt,10pt) and (35pt,-10pt).. (30pt,-30pt);
\draw[densely dashed] (30pt,30pt) .. controls (25pt,10pt) and (25pt,-10pt).. (30pt,-30pt);
\draw [-<,>=stealth] (33.7pt,-1pt) -- (33.7pt,1pt);
\draw [->,>=stealth] (15pt,30pt) -- (16pt,30pt);
\draw [-<,>=stealth] (15pt,-30pt) -- (16pt,-30pt);
\end{tikzpicture}
\qquad
\begin{tikzpicture}[scale=0.7, baseline={([yshift=-4pt]current bounding box.center)}]
\draw (0pt,30pt) -- (60pt,30pt);
\draw(0pt,-30pt) -- (60pt,-30pt);
\filldraw[fill=black] (30pt,30pt) circle (1pt)node[above]{$J$} ;
\filldraw[fill=black] (30pt,-30pt) circle (1pt) node[below]{$J$};
\draw (30pt,30pt) .. controls (35pt,10pt) and (35pt,-10pt).. (30pt,-30pt);
\draw (30pt,30pt) .. controls (25pt,10pt) and (25pt,-10pt).. (30pt,-30pt);
\draw [->,>=stealth] (26.3pt,-1pt) -- (26.3pt,1pt);
\draw [-<,>=stealth] (33.7pt,-1pt) -- (33.7pt,1pt);
\draw [->,>=stealth] (15pt,30pt) -- (16pt,30pt);
\draw [-<,>=stealth] (15pt,-30pt) -- (16pt,-30pt);
\draw [->,>=stealth] (45pt,30pt) -- (46pt,30pt);
\draw [-<,>=stealth] (45pt,-30pt) -- (46pt,-30pt);
\end{tikzpicture}
\end{align}
Note that we have reduced the vertices in Fig.~\ref{fig:vertices} to points. The indices structure for the first two diagrams are simple: $i$ must go to $j$:
\begin{align}
\begin{tikzpicture}[scale=0.7, baseline={([yshift=-4pt]current bounding box.center)}]
\node[left] at(0pt,30pt) {$i,p$};
\node[left] at(0pt,-30pt) {$i,p$};
\node[right] at(60pt,30pt) {$j,p$};
\node[right] at(60pt,-30pt) {$j,p$};
\node[left] at(30pt,0pt) {$i,\alpha$};
\node[right] at(30pt,0pt) {$j,\alpha$};
\draw[densely dashed] (0pt,30pt) -- (60pt,30pt);
\draw[densely dashed](0pt,-30pt) -- (60pt,-30pt);
\filldraw[fill=black] (30pt,30pt) circle (1pt)node[above]{$t$} ;
\filldraw[fill=black] (30pt,-30pt) circle (1pt) node[below]{$t$};
\draw (30pt,30pt) .. controls (35pt,10pt) and (35pt,-10pt).. (30pt,-30pt);
\draw (30pt,30pt) .. controls (25pt,10pt) and (25pt,-10pt).. (30pt,-30pt);
\draw [->,>=stealth] (26.3pt,-1pt) -- (26.3pt,1pt);
\draw [-<,>=stealth] (33.7pt,-1pt) -- (33.7pt,1pt);
\end{tikzpicture}
\qquad 
\begin{tikzpicture}[scale=0.7, baseline={([yshift=-4pt]current bounding box.center)}]
\node[left] at(0pt,30pt) {$i,\alpha$};
\node[left] at(0pt,-30pt) {$i,\alpha$};
\node[right] at(60pt,30pt) {$j,\alpha$};
\node[right] at(60pt,-30pt) {$j,\alpha$};
\node[left] at(30pt,0pt) {$i,p$};
\node[right] at(30pt,0pt) {$j,p$};
\draw (0pt,30pt) -- (60pt,30pt);
\draw(0pt,-30pt) -- (60pt,-30pt);
\filldraw[fill=black] (30pt,30pt) circle (1pt)node[above]{$t$} ;
\filldraw[fill=black] (30pt,-30pt) circle (1pt) node[below]{$t$};
\draw[densely dashed] (30pt,30pt) .. controls (35pt,10pt) and (35pt,-10pt).. (30pt,-30pt);
\draw[densely dashed] (30pt,30pt) .. controls (25pt,10pt) and (25pt,-10pt).. (30pt,-30pt);
\draw [->,>=stealth] (15pt,30pt) -- (16pt,30pt);
\draw [-<,>=stealth] (15pt,-30pt) -- (16pt,-30pt);
\draw [->,>=stealth] (45pt,30pt) -- (46pt,30pt);
\draw [-<,>=stealth] (45pt,-30pt) -- (46pt,-30pt);
\end{tikzpicture}
\end{align}
However the rest two ladders both have two choices:
\begin{align}
\begin{tikzpicture}[scale=0.7, baseline={([yshift=-4pt]current bounding box.center)}]
\node[left] at(0pt,30pt) {$i,\alpha$};
\node[left] at(0pt,-30pt) {$i,\alpha$};
\node[right] at(60pt,30pt) {$j,p$};
\node[right] at(60pt,-30pt) {$j,p$};
\node[left] at(30pt,0pt) {$i,p$};
\node[right] at(30pt,0pt) {$j,\alpha$};
\draw (0pt,30pt) -- (30pt,30pt);
\draw(0pt,-30pt) -- (30pt,-30pt);
\draw[densely dashed] (60pt,30pt) -- (30pt,30pt);
\draw[densely dashed] (60pt,-30pt) -- (30pt,-30pt);
\filldraw[fill=black] (30pt,30pt) circle (1pt)node[above]{$t$} ;
\filldraw[fill=black] (30pt,-30pt) circle (1pt) node[below]{$t$};
\draw (30pt,30pt) .. controls (35pt,10pt) and (35pt,-10pt).. (30pt,-30pt);
\draw[densely dashed] (30pt,30pt) .. controls (25pt,10pt) and (25pt,-10pt).. (30pt,-30pt);
\draw [-<,>=stealth] (33.7pt,-1pt) -- (33.7pt,1pt);
\draw [->,>=stealth] (15pt,30pt) -- (16pt,30pt);
\draw [-<,>=stealth] (15pt,-30pt) -- (16pt,-30pt);
\end{tikzpicture}+
\begin{tikzpicture}[scale=0.7, baseline={([yshift=-4pt]current bounding box.center)}]
\node[left] at(0pt,30pt) {$i,\alpha$};
\node[left] at(0pt,-30pt) {$i,\alpha$};
\node[right] at(60pt,30pt) {$i,p$};
\node[right] at(60pt,-30pt) {$i,p$};
\node[left] at(30pt,0pt) {$j,p$};
\node[right] at(30pt,0pt) {$j,\alpha$};
\draw (0pt,30pt) -- (30pt,30pt);
\draw(0pt,-30pt) -- (30pt,-30pt);
\draw[densely dashed] (60pt,30pt) -- (30pt,30pt);
\draw[densely dashed] (60pt,-30pt) -- (30pt,-30pt);
\filldraw[fill=black] (30pt,30pt) circle (1pt)node[above]{$t$} ;
\filldraw[fill=black] (30pt,-30pt) circle (1pt) node[below]{$t$};
\draw (30pt,30pt) .. controls (35pt,10pt) and (35pt,-10pt).. (30pt,-30pt);
\draw[densely dashed] (30pt,30pt) .. controls (25pt,10pt) and (25pt,-10pt).. (30pt,-30pt);
\draw [-<,>=stealth] (33.7pt,-1pt) -- (33.7pt,1pt);
\draw [->,>=stealth] (15pt,30pt) -- (16pt,30pt);
\draw [-<,>=stealth] (15pt,-30pt) -- (16pt,-30pt);
\end{tikzpicture} \quad ; \quad 
\begin{tikzpicture}[scale=0.7, baseline={([yshift=-4pt]current bounding box.center)}]
\node[left] at(0pt,30pt) {$i,\alpha$};
\node[left] at(0pt,-30pt) {$i,\alpha$};
\node[right] at(60pt,30pt) {$i,\beta$};
\node[right] at(60pt,-30pt) {$i,\beta$};
\node[left] at(30pt,0pt) {$j,\beta$};
\node[right] at(30pt,0pt) {$j,\alpha$};
\draw (0pt,30pt) -- (60pt,30pt);
\draw(0pt,-30pt) -- (60pt,-30pt);
\filldraw[fill=black] (30pt,30pt) circle (1pt)node[above]{$J$} ;
\filldraw[fill=black] (30pt,-30pt) circle (1pt) node[below]{$J$};
\draw (30pt,30pt) .. controls (35pt,10pt) and (35pt,-10pt).. (30pt,-30pt);
\draw (30pt,30pt) .. controls (25pt,10pt) and (25pt,-10pt).. (30pt,-30pt);
\draw [->,>=stealth] (26.3pt,-1pt) -- (26.3pt,1pt);
\draw [-<,>=stealth] (33.7pt,-1pt) -- (33.7pt,1pt);
\draw [->,>=stealth] (15pt,30pt) -- (16pt,30pt);
\draw [-<,>=stealth] (15pt,-30pt) -- (16pt,-30pt);
\draw [->,>=stealth] (45pt,30pt) -- (46pt,30pt);
\draw [-<,>=stealth] (45pt,-30pt) -- (46pt,-30pt);
\end{tikzpicture}
+
\begin{tikzpicture}[scale=0.7, baseline={([yshift=-4pt]current bounding box.center)}]
\node[left] at(0pt,30pt) {$i,\alpha$};
\node[left] at(0pt,-30pt) {$i,\alpha$};
\node[right] at(60pt,30pt) {$j,\alpha$};
\node[right] at(60pt,-30pt) {$j,\alpha$};
\node[left] at(30pt,0pt) {$j,\beta$};
\node[right] at(30pt,0pt) {$i,\beta$};
\draw (0pt,30pt) -- (60pt,30pt);
\draw(0pt,-30pt) -- (60pt,-30pt);
\filldraw[fill=black] (30pt,30pt) circle (1pt)node[above]{$J$} ;
\filldraw[fill=black] (30pt,-30pt) circle (1pt) node[below]{$J$};
\draw (30pt,30pt) .. controls (35pt,10pt) and (35pt,-10pt).. (30pt,-30pt);
\draw (30pt,30pt) .. controls (25pt,10pt) and (25pt,-10pt).. (30pt,-30pt);
\draw [->,>=stealth] (26.3pt,-1pt) -- (26.3pt,1pt);
\draw [-<,>=stealth] (33.7pt,-1pt) -- (33.7pt,1pt);
\draw [->,>=stealth] (15pt,30pt) -- (16pt,30pt);
\draw [-<,>=stealth] (15pt,-30pt) -- (16pt,-30pt);
\draw [->,>=stealth] (45pt,30pt) -- (46pt,30pt);
\draw [-<,>=stealth] (45pt,-30pt) -- (46pt,-30pt);
\end{tikzpicture}
\end{align}
Therefore we need to consider the extra counting when we have such ladder. Another subtlety arises from $k=M'/M$ factor. Each closed solid loop contributes $M$ and dashed loop contributes $M'$. 
Since we are interested in the site-uniform fluctuations, and mainly calculate $\Pi_1=\frac{1}{N}\sum_{ij} \Pi_1^{ij} $ which washes out the index structure and only keeps the multiplicative factors for the diagram counting, we will not label the indices for the lines in the following equations. 

As we have defined in the main text, the $\Pi_0$ is given by
single bubble: 
\begin{align}
\Pi_0:\quad 
\begin{tikzpicture}[scale=0.7, baseline={([yshift=-4pt]current bounding box.center)}]
\filldraw[fill=black] (-60pt,0pt) circle (1pt) node[left] {$\tau_1$};
\filldraw[fill=black] (60pt,0pt) circle (1pt) node[right] {$\tau_2$};
\draw[densely dashed] (-60pt,0pt) ..  controls (-15pt,15pt) and (15pt,15pt).. (60pt,0pt);
\draw[densely dashed] (-60pt,0pt) ..  controls (-15pt,-15pt) and (15pt,-15pt).. (60pt,0pt);
\end{tikzpicture}
= \chi(\tau_1-\tau_2)^2
\label{fig:Pi0}
\end{align}
$\Pi_1$ represents the ladder diagrams and is given by the sum of infinite series:
\begin{align}
\Pi_1:\quad
\begin{tikzpicture}[scale=0.7,baseline={([yshift=-4pt]current bounding box.center)}]
\filldraw[fill=black] (-40pt,0pt) circle (1pt) ;
\filldraw[fill=black] (40pt,0pt) circle (1pt) ;
\draw[densely dashed] (-40pt,0pt) -- (0pt,30pt) ;
\draw[densely dashed] (40pt,0pt) -- (0pt,30pt);
\draw[densely dashed] (-40pt,0pt) -- (0pt,-30pt) ;
\draw[densely dashed] (40pt,0pt) -- (0pt,-30pt);
\filldraw[fill=black] (0pt,30pt) circle (1pt) node[above]{$t$};
\filldraw[fill=black] (0pt,-30pt) circle (1pt) node[below]{$t$};
\draw (0pt,30pt) .. controls (5pt,10pt) and (5pt,-10pt).. (0pt,-30pt);
\draw (0pt,30pt) .. controls (-5pt,10pt) and (-5pt,-10pt).. (0pt,-30pt);
\draw [->,>=stealth] (-3.7pt,-1pt) -- (-3.7pt,1pt);
\draw [-<,>=stealth] (3.7pt,-1pt) -- (3.7pt,1pt);
\end{tikzpicture}
\quad
+
\quad
\begin{tikzpicture}[scale=0.7,baseline={([yshift=-4pt]current bounding box.center)}]
\filldraw[fill=black] (-40pt,0pt) circle (1pt) ;
\filldraw[fill=black] (70pt,0pt) circle (1pt)  ;
\draw[densely dashed] (-40pt,0pt) -- (0pt,30pt) ;
\draw[densely dashed] (70pt,0pt) -- (30pt,30pt);
\draw[densely dashed] (-40pt,0pt) -- (0pt,-30pt) ;
\draw[densely dashed] (70pt,0pt) -- (30pt,-30pt);
\filldraw[fill=black] (0pt,30pt) circle (1pt) node[above]{$t$};
\filldraw[fill=black] (0pt,-30pt) circle (1pt) node[below]{$t$};
\draw (0pt,30pt) .. controls (5pt,10pt) and (5pt,-10pt).. (0pt,-30pt);
\draw (0pt,30pt) .. controls (-5pt,10pt) and (-5pt,-10pt).. (0pt,-30pt);
\draw[densely dashed] (0pt,30pt) -- (30pt,30pt);
\draw[densely dashed] (0pt,-30pt) -- (30pt,-30pt);
\filldraw[fill=black] (30pt,30pt) circle (1pt) node[above]{$t$};
\filldraw[fill=black] (30pt,-30pt) circle (1pt) node[below]{$t$};
\draw (30pt,30pt) .. controls (35pt,10pt) and (35pt,-10pt).. (30pt,-30pt);
\draw (30pt,30pt) .. controls (25pt,10pt) and (25pt,-10pt).. (30pt,-30pt);
\draw [->,>=stealth] (26.3pt,-1pt) -- (26.3pt,1pt);
\draw [-<,>=stealth] (33.7pt,-1pt) -- (33.7pt,1pt);
\draw [->,>=stealth] (-3.7pt,-1pt) -- (-3.7pt,1pt);
\draw [-<,>=stealth] (3.7pt,-1pt) -- (3.7pt,1pt);
\end{tikzpicture}
\quad
+
\quad
\begin{tikzpicture}[scale=0.7, baseline={([yshift=-4pt]current bounding box.center)}]
\filldraw[fill=black] (-40pt,0pt) circle (1pt) ;
\filldraw[fill=black] (70pt,0pt) circle (1pt)  ;
\draw[densely dashed] (-40pt,0pt) -- (0pt,30pt) ;
\draw[densely dashed] (70pt,0pt) -- (30pt,30pt);
\draw[densely dashed] (-40pt,0pt) -- (0pt,-30pt) ;
\draw[densely dashed] (70pt,0pt) -- (30pt,-30pt);
\filldraw[fill=black] (0pt,30pt) circle (1pt) node[above]{$t$};
\filldraw[fill=black] (0pt,-30pt) circle (1pt) node[below]{$t$};
\draw[densely dashed] (0pt,30pt) .. controls (5pt,10pt) and (5pt,-10pt).. (0pt,-30pt);
\draw (0pt,30pt) .. controls (-5pt,10pt) and (-5pt,-10pt).. (0pt,-30pt);
\draw (0pt,30pt) -- (30pt,30pt);
\draw (0pt,-30pt) -- (30pt,-30pt);
\filldraw[fill=black] (30pt,30pt) circle (1pt) node[above]{$t$};
\filldraw[fill=black] (30pt,-30pt) circle (1pt) node[below]{$t$};
\draw (30pt,30pt) .. controls (35pt,10pt) and (35pt,-10pt).. (30pt,-30pt);
\draw[densely dashed] (30pt,30pt) .. controls (25pt,10pt) and (25pt,-10pt).. (30pt,-30pt);
\draw [->,>=stealth] (-3.7pt,-1pt) -- (-3.7pt,1pt);
\draw [-<,>=stealth] (33.7pt,-1pt) -- (33.7pt,1pt);
\draw [->,>=stealth] (15pt,30pt) -- (16pt,30pt);
\draw [-<,>=stealth] (15pt,-30pt) -- (16pt,-30pt);
\end{tikzpicture}
\quad
\ldots
\end{align}
When we have longer ladders, it could also involve four fermion vertices, e.g. we would have both $t$-$t$ ladder and $J$-$J$ ladder and they have same form of propagators (upto a constant factor) running in the ladder since both $\chi(\tau_1-\tau_2)^2$ and $G(\tau_1-\tau_2)^2$ are proportional to $|\tau_1-\tau_2|^{-1}$ for $\Delta_b=\Delta_f=\frac{1}{4}$.
\begin{align}
\begin{tikzpicture}[scale=0.7, baseline={([yshift=-4pt]current bounding box.center)}]
\filldraw[fill=black] (-40pt,0pt) circle (1pt) ;
\draw[densely dashed] (-40pt,0pt) -- (0pt,30pt) ;
\draw[densely dashed] (-40pt,0pt) -- (0pt,-30pt) ;
\filldraw[fill=black] (0pt,30pt) circle (1pt) ;
\filldraw[fill=black] (0pt,-30pt) circle (1pt) ;
\draw[densely dashed] (0pt,30pt) .. controls (5pt,10pt) and (5pt,-10pt).. (0pt,-30pt);
\draw (0pt,30pt) .. controls (-5pt,10pt) and (-5pt,-10pt).. (0pt,-30pt);
\draw (0pt,30pt) -- (60pt,30pt);
\draw(0pt,-30pt) -- (60pt,-30pt);
\filldraw[fill=black] (30pt,30pt) circle (1pt)node[above]{$t$} ;
\filldraw[fill=black] (30pt,-30pt) circle (1pt) node[below]{$t$};
\draw[densely dashed] (30pt,30pt) .. controls (35pt,10pt) and (35pt,-10pt).. (30pt,-30pt);
\draw[densely dashed] (30pt,30pt) .. controls (25pt,10pt) and (25pt,-10pt).. (30pt,-30pt);
\filldraw[fill=black] (60pt,30pt) circle (1pt) ;
\filldraw[fill=black] (60pt,-30pt) circle (1pt) ;
\draw (60pt,30pt) .. controls (65pt,10pt) and (65pt,-10pt).. (60pt,-30pt);
\draw[densely dashed] (60pt,30pt) .. controls (55pt,10pt) and (55pt,-10pt).. (60pt,-30pt);
\draw[densely dashed] (100pt,0pt) -- (60pt,30pt);
\draw[densely dashed] (100pt,0pt) -- (60pt,-30pt);
\filldraw[fill=black] (100pt,0pt) circle (1pt);
\draw [->,>=stealth] (-3.7pt,-1pt) -- (-3.7pt,1pt);
\draw [-<,>=stealth] (63.7pt,-1pt) -- (63.7pt,1pt);
\draw [->,>=stealth] (15pt,30pt) -- (16pt,30pt);
\draw [-<,>=stealth] (15pt,-30pt) -- (16pt,-30pt);
\draw [->,>=stealth] (45pt,30pt) -- (46pt,30pt);
\draw [-<,>=stealth] (45pt,-30pt) -- (46pt,-30pt);
\end{tikzpicture}
\quad
+
\quad
\begin{tikzpicture}[scale=0.7, baseline={([yshift=-4pt]current bounding box.center)}]
\filldraw[fill=black] (-40pt,0pt) circle (1pt) ;
\draw[densely dashed] (-40pt,0pt) -- (0pt,30pt) ;
\draw[densely dashed] (-40pt,0pt) -- (0pt,-30pt) ;
\filldraw[fill=black] (0pt,30pt) circle (1pt) ;
\filldraw[fill=black] (0pt,-30pt) circle (1pt) ;
\draw[densely dashed] (0pt,30pt) .. controls (5pt,10pt) and (5pt,-10pt).. (0pt,-30pt);
\draw (0pt,30pt) .. controls (-5pt,10pt) and (-5pt,-10pt).. (0pt,-30pt);
\draw (0pt,30pt) -- (60pt,30pt);
\draw(0pt,-30pt) -- (60pt,-30pt);
\filldraw[fill=black] (30pt,30pt) circle (1pt)node[above]{$J$} ;
\filldraw[fill=black] (30pt,-30pt) circle (1pt) node[below]{$J$};
\draw (30pt,30pt) .. controls (35pt,10pt) and (35pt,-10pt).. (30pt,-30pt);
\draw (30pt,30pt) .. controls (25pt,10pt) and (25pt,-10pt).. (30pt,-30pt);
\filldraw[fill=black] (60pt,30pt) circle (1pt) ;
\filldraw[fill=black] (60pt,-30pt) circle (1pt) ;
\draw (60pt,30pt) .. controls (65pt,10pt) and (65pt,-10pt).. (60pt,-30pt);
\draw[densely dashed] (60pt,30pt) .. controls (55pt,10pt) and (55pt,-10pt).. (60pt,-30pt);
\draw[densely dashed] (100pt,0pt) -- (60pt,30pt);
\draw[densely dashed] (100pt,0pt) -- (60pt,-30pt);
\filldraw[fill=black] (100pt,0pt) circle (1pt);
\draw [->,>=stealth] (-3.7pt,-1pt) -- (-3.7pt,1pt);
\draw [-<,>=stealth] (33.7pt,-1pt) -- (33.7pt,1pt);
\draw [->,>=stealth] (26.3pt,-1pt) -- (26.3pt,1pt);
\draw [-<,>=stealth] (63.7pt,-1pt) -- (63.7pt,1pt);
\draw [->,>=stealth] (15pt,30pt) -- (16pt,30pt);
\draw [-<,>=stealth] (15pt,-30pt) -- (16pt,-30pt);
\draw [->,>=stealth] (45pt,30pt) -- (46pt,30pt);
\draw [-<,>=stealth] (45pt,-30pt) -- (46pt,-30pt);
\end{tikzpicture}
\quad+ \quad
\begin{tikzpicture}[scale=0.7, baseline={([yshift=-4pt]current bounding box.center)}]
\filldraw[fill=black] (-40pt,0pt) circle (1pt) ;
\draw[densely dashed] (-40pt,0pt) -- (0pt,30pt) ;
\draw[densely dashed] (-40pt,0pt) -- (0pt,-30pt) ;
\filldraw[fill=black] (0pt,30pt) circle (1pt) ;
\filldraw[fill=black] (0pt,-30pt) circle (1pt) ;
\draw[densely dashed] (0pt,30pt) .. controls (5pt,10pt) and (5pt,-10pt).. (0pt,-30pt);
\draw (0pt,30pt) .. controls (-5pt,10pt) and (-5pt,-10pt).. (0pt,-30pt);
\draw (0pt,30pt) -- (60pt,30pt);
\draw(0pt,-30pt) -- (60pt,-30pt);
\filldraw[fill=black] (30pt,30pt) circle (1pt)node[above]{$J$} ;
\filldraw[fill=black] (30pt,-30pt) circle (1pt) node[below]{$J$};
\draw (30pt,30pt) .. controls (35pt,10pt) and (35pt,-10pt).. (30pt,-30pt);
\draw (30pt,30pt) .. controls (25pt,10pt) and (25pt,-10pt).. (30pt,-30pt);
\filldraw[fill=black] (60pt,30pt) circle (1pt) ;
\filldraw[fill=black] (60pt,-30pt) circle (1pt) ;
\draw (60pt,30pt) .. controls (65pt,10pt) and (65pt,-10pt).. (60pt,-30pt);
\draw[densely dashed] (60pt,30pt) .. controls (55pt,10pt) and (55pt,-10pt).. (60pt,-30pt);
\draw[densely dashed] (100pt,0pt) -- (60pt,30pt);
\draw[densely dashed] (100pt,0pt) -- (60pt,-30pt);
\filldraw[fill=black] (100pt,0pt) circle (1pt);
\draw [->,>=stealth] (-3.7pt,-1pt) -- (-3.7pt,1pt);
\draw [-<,>=stealth] (33.7pt,-1pt) -- (33.7pt,1pt);
\draw [-<,>=stealth] (26.3pt,-1pt) -- (26.3pt,1pt);
\draw [->,>=stealth] (63.7pt,-1pt) -- (63.7pt,1pt);
\draw [->,>=stealth] (15pt,30pt) -- (16pt,30pt);
\draw [-<,>=stealth] (15pt,-30pt) -- (16pt,-30pt);
\draw [-<,>=stealth] (45pt,30pt) -- (46pt,30pt);
\draw [->,>=stealth] (45pt,-30pt) -- (46pt,-30pt);
\end{tikzpicture}
\end{align}


\subsection{Recurrence relation}
A convenient way to sum all the diagrams is to derive a recurrence relation for the diagrams with different number of ladders. Note that we are working in the case $\Delta_b=\Delta_f=1/4$, therefore the difference (upto constant factor) between boson lines and fermion lines comes from the $\sgn(\tau)$ factor, which will be represented by arrows in the diagrams.

We first derive the elementary reduction. 
For convenience, we will use black line to represent $|\tau_1-\tau_2|^0=\const$, red line to represent $|\tau_1-\tau_2|^{-1}$, blue line to represent $|\tau_1-\tau_2|^{-1/2}$ and use arrow from vertex $j$ to vertex $k$ to represent $-\sgn(\tau_j-\tau_k)$.

For fermion block, we can reduce a block consist of four fermion lines with scaling dimension $\Delta= 1/4$:
\begin{align}
\begin{tikzpicture}[baseline={([yshift=-4pt]current bounding box.center)}]
\filldraw[fill=black] (-40pt,0pt) circle (1pt) ;
\draw[blue] (-40pt,0pt) -- (0pt,30pt) ;
\draw[blue] (-40pt,0pt) -- (0pt,-30pt) ;
\filldraw[fill=black] (0pt,30pt) circle (1pt) ;
\filldraw[fill=black] (0pt,-30pt) circle (1pt) ;
\draw[red,->,>=stealth] (0pt,-30pt) -- (0pt,0pt);
\draw[red] (0pt,30pt) -- (0pt,-30pt);
\draw[blue] (0pt,30pt) -- (50pt,30pt);
\draw[blue,->,>=stealth] (0pt,30pt) -- (25pt,30pt);
\draw[blue] (0pt,-30pt) -- (50pt,-30pt);
\draw[blue,-<,>=stealth] (0pt,-30pt) -- (25pt,-30pt);
\filldraw[fill=black] (50pt,30pt) circle (1pt);
\filldraw[fill=black] (50pt,-30pt) circle (1pt);
\draw[blue] (50pt,30pt)--(50pt,-30pt);
\draw[blue,->,>=stealth] (50pt,30pt)--(50pt,0pt);
\end{tikzpicture}
= 
A_1 \quad
\begin{tikzpicture}[baseline={([yshift=-4pt]current bounding box.center)}]
\filldraw[fill=black] (-40pt,0pt) circle (1pt) ;
\draw (-40pt,0pt) -- (50pt,30pt);
\draw[->,>=stealth] (-40pt,0pt) -- (5pt,15pt);
\draw[red,-<,>=stealth] (-40pt,0pt) -- (-20pt,-15pt);
\draw[red] (-40pt,0pt) -- (0pt,-30pt);
\filldraw[fill=black] (0pt,-30pt) circle (1pt) ;
\draw[blue] (50pt,30pt) -- (0pt,-30pt);
\draw[blue] (0pt,-30pt) -- (50pt,-30pt);
\draw[blue,-<,>=stealth] (0pt,-30pt) -- (25pt,-30pt);
\filldraw[fill=black] (50pt,30pt) circle (1pt);
\filldraw[fill=black] (50pt,-30pt) circle (1pt);
\draw[blue] (50pt,30pt)--(50pt,-30pt);
\draw[blue,->,>=stealth] (50pt,30pt)--(50pt,0pt);
\end{tikzpicture}
=
A_1 A_2 \quad
\begin{tikzpicture}[baseline={([yshift=-4pt]current bounding box.center)}]
\filldraw[fill=black] (-40pt,0pt) circle (1pt) ;
\draw[blue] (-40pt,0pt) -- (50pt,30pt);
\draw[blue] (-40pt,0pt) -- (50pt,-30pt);
\filldraw[fill=black] (50pt,30pt) circle (1pt);
\filldraw[fill=black] (50pt,-30pt) circle (1pt);
\draw[blue] (50pt,30pt)--(50pt,-30pt);
\end{tikzpicture}
\end{align}
In each step we integrate over one time (represented by a point) using a star-triangle identity\cite{Symanzik:1972wj} (also see section 2.2.3 of \cite{kitaev2017} for a reference and similar application).
$A_1$ and $A_2$ are coefficients deduced from star-triangle identity:
\begin{align}
A_1= \frac{4}{\pi} \Gamma \left( \frac{1}{2} \right) ^2 \Gamma(0) \sin \frac{\pi}{4} \cos \frac{\pi}{4}  \cos \frac{\pi}{2} =2, \quad
A_2= A_1=2
\end{align}

The bosonic block needs further treatment: the boson $\frac{1}{\tau}$ line will induce a $\Gamma(0)$ divergence while applying the star-triangle identity and need to be regulated. We can consider a dimension regularization and shift the left three lines in a way that we can still apply the star-triangle identity (the total scaling dimension at a vertex is 1):
\begin{align}
\begin{tikzpicture}[baseline={([yshift=-4pt]current bounding box.center)}]
\filldraw[fill=black] (-40pt,0pt) circle (1pt) ;
\draw[blue] (-40pt,0pt) -- (0pt,30pt) ;
\draw[blue] (-40pt,0pt) -- (0pt,-30pt) ;
\filldraw[fill=black] (0pt,30pt) circle (1pt) ;
\filldraw[fill=black] (0pt,-30pt) circle (1pt) ;
\draw[red] (0pt,30pt) -- (0pt,-30pt);
\draw[blue] (0pt,30pt) -- (50pt,30pt);
\draw[blue] (0pt,-30pt) -- (50pt,-30pt);
\filldraw[fill=black] (50pt,30pt) circle (1pt);
\filldraw[fill=black] (50pt,-30pt) circle (1pt);
\draw[blue] (50pt,30pt)--(50pt,-30pt);
\node at (-30pt,20pt) {\scriptsize $\frac{1}{4}+\frac{\epsilon}{2}$};
\node at (-30pt,-20pt) {\scriptsize $\frac{1}{4}+\frac{\epsilon}{2}$};
\node at (12pt,0pt) {\scriptsize $\frac{1}{2}-\frac{\epsilon}{2}$};
\node at (20pt,40pt) {\scriptsize $\frac{1}{4}$};
\node at (20pt,-40pt) {\scriptsize $\frac{1}{4}$};
\node at (45pt,0pt) {\scriptsize $\frac{1}{4}$};
\end{tikzpicture}
= 
A'_1 \quad
\begin{tikzpicture}[baseline={([yshift=-4pt]current bounding box.center)}]
\filldraw[fill=black] (-40pt,0pt) circle (1pt) ;
\draw (-40pt,0pt) -- (50pt,30pt);
\draw[red] (-40pt,0pt) -- (0pt,-30pt);
\filldraw[fill=black] (0pt,-30pt) circle (1pt) ;
\draw[blue] (50pt,30pt) -- (0pt,-30pt);
\draw[blue] (0pt,-30pt) -- (50pt,-30pt);
\filldraw[fill=black] (50pt,30pt) circle (1pt);
\filldraw[fill=black] (50pt,-30pt) circle (1pt);
\draw[blue] (50pt,30pt)--(50pt,-30pt);
\node at (0pt,20pt) {\scriptsize $\frac{\epsilon}{2}$};
\node at (-30pt,-20pt) {\scriptsize $\frac{1}{2}+\frac{\epsilon}{2}$};
\node at (12pt,0pt) {\scriptsize $\frac{1}{4}-\frac{\epsilon}{2}$};
\node at (20pt,-40pt) {\scriptsize $\frac{1}{4}$};
\node at (45pt,0pt) {\scriptsize $\frac{1}{4}$};
\end{tikzpicture}
=
A'_1 A'_2 \quad
\begin{tikzpicture}[baseline={([yshift=-4pt]current bounding box.center)}]
\filldraw[fill=black] (-40pt,0pt) circle (1pt) ;
\draw[blue] (-40pt,0pt) -- (50pt,30pt);
\draw[blue] (-40pt,0pt) -- (50pt,-30pt);
\filldraw[fill=black] (50pt,30pt) circle (1pt);
\filldraw[fill=black] (50pt,-30pt) circle (1pt);
\draw[blue] (50pt,30pt)--(50pt,-30pt);
\node at (-5pt,20pt) {\scriptsize $\frac{1}{4}+ \frac{\epsilon}{2}$};
\node at (-5pt,-20pt) {\scriptsize $\frac{1}{4}+ \frac{\epsilon}{2}$};
\node at (36pt,0pt) {\scriptsize $\frac{1}{4}-\frac{\epsilon}{2}$};
\end{tikzpicture}
\end{align}

\begin{align}
A'_1= \frac{4}{\pi} \Gamma \left( \frac{1}{2} \right) ^2 \Gamma(\epsilon) \sin^2 \frac{\pi}{4} \sin \frac{\pi}{2} =\frac{2}{\epsilon}, \quad
A'_2= \frac{4}{\pi} \Gamma \left( \frac{1}{2} \right) ^2 \Gamma(-\epsilon) \sin^2 \frac{\pi}{4} \sin \frac{\pi}{2} =-\frac{2}{\epsilon}
\end{align}
The small dimension shift $\epsilon$ can be related to the cut-off scale $\Lambda$ using following argument: the divergence in the above diagram for $\frac{1}{|\tau|}$ arises from the fourier transformation, where we need to introduce a cut-off $\Lambda$, $\int_{\Lambda^{-1}}^{\infty} d\tau e^{i\omega t}/|\tau| \sim \log \frac{\Lambda}{\omega}$. We can estimate the inverse fourier transformaiton using the following approximation (we put a time scale $|\tau_{12}|$, the time sapration of two $\lambda$ field to make the expression dimensionally sensible)
\begin{align}
\log \frac{\Lambda |\tau_{12}|}{|\omega \tau_{12}|} = \log \Lambda |\tau_{12}| \left( 1 - \frac{\log |\omega \tau_{12}|}{\log \Lambda |\tau_{12}|} \right)  \approx \log \Lambda |\tau_{12}| \cdot |\omega \tau_{12}|^{-\frac{1}{\log \Lambda |\tau_{12}|}} \xrightarrow{\text{F.T.}} \frac{1}{{|\tau|}} \cdot \left\vert \frac{\tau}{\tau_{12}} \right\vert^{\frac{1}{\log \Lambda |\tau_{12}|}} 
\end{align}
Therefore the cut-off amounts to shifting the scaling dimension $\frac{1}{2}$ to $\frac{1}{2}- \frac{\epsilon}{2}$ where $\epsilon= \frac{1}{\log \Lambda\tau_{12}}$.

Having these two elementary reduction we are able to derive recurrence relations. First, we notice the diagram counting has significant difference for bosonic box and fermionic box, so we define two different types of ladder:
\begin{enumerate}
\item $\Pi_{B,n}$: sum of all $n$ ladders diagrams that start with a bosonic box, i.e. the first (left) vertical ladder is arrowless;

\item  $\Pi_{F,n}$: sum of all $n$ ladders diagrams that start with a fermionic box, i.e. the first vertical ladder has an arrow on it.
\end{enumerate}
Now we consider how to get $\Pi_{B/F,n+1}$ from $\Pi_{B/F,n}$. We can add a bosonic box to $n$ ladders to get $\Pi_{B,n+1}$ straightforwardly:
\begin{align}
\Pi_{B,n+1}={A'_1A'_2 \cdot \frac{t^2}{NM} N M \frac{C^2 F^2}{J^4}}\cdot \left(  \Pi_{B,n}+\Pi_{F,n} \right) =a \Pi_{B,n}+a\Pi_{F,n} ,\quad a=- \frac{1}{2\pi \epsilon^2}
\end{align} 
where we use Eq.~(\ref{e2}) for the relations of coefficients $C,F$ to evaluate $a$. 
The fermionic $\Pi_{F,n+1}$ is more involved: if we add a fermionic block to $\Pi_{B,n}$, we have:
\begin{align}
-A_1 A_2 \cdot \frac{8t^2}{NM} NM' \frac{C^2 F^2}{J^4}\cdot \Pi_{B,n} =b \Pi_{B,n}, \quad b= -\frac{4k}{\pi}
\end{align}
Now if we add an extra fermionic box adding to $\Pi_{F,n}$, there are two possibilities: one can add a $t$-$t$ vertex or $J$-$J$ vertex, the contribution in total is
\begin{align}
-A_1A_2 \cdot \left(  \frac{t^2}{NM} NM' \frac{C^2F^2}{J^4} + \frac{3}{2} 
\frac{J^2}{NM} NM \frac{F^4}{J^2}
 \right) \cdot \Pi_{F,n} = c  \Pi_{F,n}, \quad c=- \frac{6-k}{4\pi}
\end{align}
Therefore the two recurrence relations are:
\begin{align}
\Pi_{B,n+1}&= a \Pi_{B,n}+a \Pi_{F,n} , \quad
\Pi_{F,n+1}= b \Pi_{B,n}+ c \Pi_{F,n}  
\end{align}
And the boundary condition is $\Pi_{F,1}=0$ and $\Pi_{B,1}= \frac{1}{4\pi \epsilon^2  }  \Pi_0$ which is shown below and evaluated using star-triangle identity again, we add the $\frac{\epsilon}{2}$ dimension shift to the right corner boson line to make the total dimension for the whole diagram unchanged. 
\begin{align}
\Pi_{B,1}=
\begin{tikzpicture}[baseline={([yshift=-4pt]current bounding box.center)}]
\draw[densely dashed] (-40pt,0pt) -- (0pt,30pt) ;
\draw[densely dashed] (40pt,0pt) -- (0pt,30pt);
\draw[densely dashed] (-40pt,0pt) -- (0pt,-30pt) ;
\draw[densely dashed] (40pt,0pt) -- (0pt,-30pt);
\filldraw[fill=black] (0pt,30pt) circle (1pt) node[above]{$t$};
\filldraw[fill=black] (0pt,-30pt) circle (1pt) node[below]{$t$};
\draw (0pt,30pt) .. controls (5pt,10pt) and (5pt,-10pt).. (0pt,-30pt);
\node at (-30pt,20pt) {\scriptsize $\frac{1}{4}+\frac{\epsilon}{2}$};
\node at (-30pt,-20pt) {\scriptsize $\frac{1}{4}+\frac{\epsilon}{2}$};
\node at (20pt,0pt) {\scriptsize $\frac{1}{2}-\frac{\epsilon}{2}$};
\node at (30pt,20pt) {\scriptsize $\frac{1}{4}$};
\node at (30pt,-20pt) {\scriptsize $\frac{1}{4}-\frac{\epsilon}{2}$};
\draw (0pt,30pt) .. controls (-5pt,10pt) and (-5pt,-10pt).. (0pt,-30pt);
\draw [->,>=stealth] (-3.7pt,-1pt) -- (-3.7pt,1pt);
\draw [-<,>=stealth] (3.7pt,-1pt) -- (3.7pt,1pt);
\end{tikzpicture}
=  \frac{1}{4\pi \epsilon^2  }  \Pi_0
\end{align}
This set of recurrence relation can be treated as equation:
\begin{align}
\begin{pmatrix}
\Pi_{B,n+1} \\
\Pi_{F,n+1}
\end{pmatrix} = 
\begin{pmatrix}
a & a \\
b & c
\end{pmatrix}
\begin{pmatrix}
\Pi_{B,n} \\
\Pi_{F,n}
\end{pmatrix}, \quad \begin{pmatrix}
\Pi_{B,1} \\
\Pi_{F,1}
\end{pmatrix} = 
\begin{pmatrix}
\frac{\Pi_0}{4\pi \epsilon^2} \\
0
\end{pmatrix}
\end{align}
Then the total sum can be expressed as:
\begin{align}
    \tilde{\Pi}_1= 
    \begin{pmatrix}
1, & 1
\end{pmatrix} \cdot
    \left(1-
\begin{pmatrix}
a & a \\
b & c
\end{pmatrix} \right)^{-1}
\cdot
\begin{pmatrix}
\Pi_{B,1} \\
\Pi_{F,1}
\end{pmatrix} =
\frac{1}{2+4\pi \epsilon^2} \Pi_0
\end{align}
There is a final subtlety that for the diagrams with fermionic box, there is an overall up-down flip with arrow reversing $\ZZ_2$ symmetry for the diagrams, which means we have double counted every diagram except the those with purely bosonic box in the above procedure. The total sum of such ladders is given by:
\begin{align}
    \tilde{\Pi}_{B}= (1+a+a^2+\ldots) \Pi_{B,1}  = \frac{1}{2+4\pi \epsilon^2} \Pi_0\,.
\end{align}
Therefore the final result is
\begin{align}
    \Pi_1(\tau_{12}) =
    \frac{1}{2} \tilde{\Pi}_1 + \frac{1}{2} \tilde{\Pi}_B \approx 
    \frac{1}{2} \left(
    1-\frac{2\pi}{(\log \Lambda |\tau_{12}|)^2}
    \right) \Pi_0(\tau_{12})\,,
    \label{eq:Pi1}
\end{align}
where 
we expand the result for small $\epsilon=\frac{1}{\log \Lambda |\tau_{12}|}$ and only show the leading terms.

\bibliography{z2syk.bib}

\end{document}